\documentclass[aps,prb,10pt,amsmath,amssymb,twocolumn,footinbib,superscriptaddress]{revtex4-1}

\usepackage{color, braket}
\usepackage[large,bf]{subfigure}     

\usepackage[breaklinks]{hyperref}

\newcommand{\surfext}{jpeg}

\definecolor{purple}{rgb}{0.5,0,0.5}
\definecolor{dkgreen}{rgb}{0,0.5,0}
\definecolor{orange}{rgb}{1,0.5,0}

\newcommand{\comment}[1]{{}}

\usepackage{bbm}

\usepackage{ifpdf}
\ifpdf
    \usepackage{psfrag}
    \usepackage[pdftex]{epsfig}
\else
    \usepackage{psfrag}
    \usepackage[dvips]{epsfig}
\fi

\begin{document}

\title{Approaching a topological phase transition in Majorana nanowires}
\date{\today}
\author{Ryan V. Mishmash}
\affiliation{Department of Physics and Institute for Quantum Information and Matter, California Institute of Technology, Pasadena, CA 91125, USA}
\affiliation{Walter Burke Institute for Theoretical Physics, California Institute of Technology, Pasadena, CA 91125, USA}
\author{David Aasen}
\affiliation{Department of Physics and Institute for Quantum Information and Matter, California Institute of Technology, Pasadena, CA 91125, USA}
\author{Andrew P. Higginbotham}
\affiliation{Department of Physics, Harvard University, Cambridge, Massachusetts, 02138, USA}
\affiliation{Center for Quantum Devices, Niels Bohr Institute, University of Copenhagen, Copenhagen, Denmark}
\author{Jason Alicea}
\affiliation{Department of Physics and Institute for Quantum Information and Matter, California Institute of Technology, Pasadena, CA 91125, USA}
\affiliation{Walter Burke Institute for Theoretical Physics, California Institute of Technology, Pasadena, CA 91125, USA}


\begin{abstract}
Recent experiments have produced mounting evidence of Majorana zero modes in nanowire-superconductor hybrids.  Signatures of an expected topological phase transition accompanying the onset of these modes nevertheless remain elusive.  We investigate a fundamental question concerning this issue: Do well-formed Majorana modes necessarily entail a sharp phase transition in these setups?  Assuming reasonable parameters, we argue that finite-size effects can dramatically smooth this putative transition into a crossover, even in systems large enough to support well-localized Majorana modes.  We propose overcoming such finite-size effects by examining the behavior of low-lying excited states through tunneling spectroscopy.  In particular, the excited-state energies exhibit characteristic field and density dependence, and scaling with system size, that expose an approaching topological phase transition.  We suggest several experiments for extracting the predicted behavior.  As a useful byproduct, the protocols also allow one to measure the wire's spin-orbit coupling directly in its superconducting environment.
\end{abstract}

\maketitle

\section{Introduction}
\label{Introduction}

Tunneling spectroscopy provides a powerful probe of topological superconductivity\cite{ReadGreen,1DwiresKitaev}.  Perhaps most notably, Majorana zero modes hosted by such systems are predicted to mediate `perfect Andreev reflection' in the asymptotic low-energy limit, generating quantized $2e^2/h$ zero-bias conductance as temperature $T \rightarrow 0$.\cite{Sengupta,ZeroBiasAnomaly3,ZeroBiasAnomaly4,ZeroBiasAnomaly6,LLjunction,BeenakkerReview2}  
In proximitized nanowires (non-quantized) zero-bias peaks were indeed observed \cite{mourik12,das12,deng12,finck12,Churchill} in the presence of a modest applied magnetic field needed to drive the system from a trivial to topological superconducting state \cite{1DwiresLutchyn,1DwiresOreg}; see also Refs.~\onlinecite{Nadj-Perge,Pawlak,BerlinAdatoms} for similar measurements on ferromagnetic atomic chains.  These experiments offer tantalizing evidence of Majorana modes and have justifiably sparked a great deal of activity.

For an \emph{infinite} proximitized nanowire a sharp second-order phase transition separates the topological and trivial states\cite{1DwiresLutchyn,1DwiresOreg,TopologicalCriticalPoints,WenBook}.  Consequently, as one ramps up the field the bulk gap closes and then reopens\footnote{It is in this sense that the phase transition is `sharp'.} in the topological phase, leaving end Majorana zero modes behind.  The collapse and revival of the bulk gap concomitant with the onset of a Majorana-induced zero-bias peak constitutes a more refined prediction that appears very difficult to mimic in alternative zero-bias-anomaly scenarios \cite{MajoranaImposter1,MajoranaImposter2,MajoranaImposter3,MajoranaImposter4,MajoranaImposter5}.  Gap revival at the putative phase transition has, however, so far proven experimentally elusive\footnote{Reference~\onlinecite{das12} reported possible observations of gap closure in some devices but not others.  However, the superconducting wire segments studied there were very short ($\sim$150nm), so that gap closure and revival should be avoided quite generally; see below.}.  Reference \onlinecite{StanescuTransition} suggested that observing this feature is difficult when tunneling into the system's ends simply because the bulk wavefunctions that become gapless at the transition tend to have little support near the boundaries.  Similar effects can appear in multi-channel wires where the `topological sub-band' may couple weakly to the lead\cite{Pientka,Prada}.

Visibility issues aside, it is useful to pose a more general question: Could existing experiments have observed the predicted bulk phase transition even as a matter of principle (say, by tunneling into the middle of the wire instead of its ends or by other proposed means \cite{TopologicalPhaseTransitionConductance,TopologicalCriticalPoints,Pientka,Pientka2,SanJose,Fregoso,Lobos,Guigou,Nozadze,KueiSun})?  
Below we argue that the previously examined superconducting wires may experience strong finite-size effects that quite drastically smooth the phase transition into a crossover.  Indeed, with reasonable assumptions we estimate that the gapless bulk modes at an infinite wire's phase transition acquire a sizable gap of order the induced pairing energy, even in the largest systems studied to date.  In this situation one should not expect to see sharp signs of a phase transition, though extended Majorana-induced zero-bias peaks can still occur provided spin-orbit coupling is sufficiently strong.  Subtler methods are then called for to verify this key aspect of the theory.  

\begin{figure*}
\centering{
\subfigure{\includegraphics[width=0.315\textwidth]{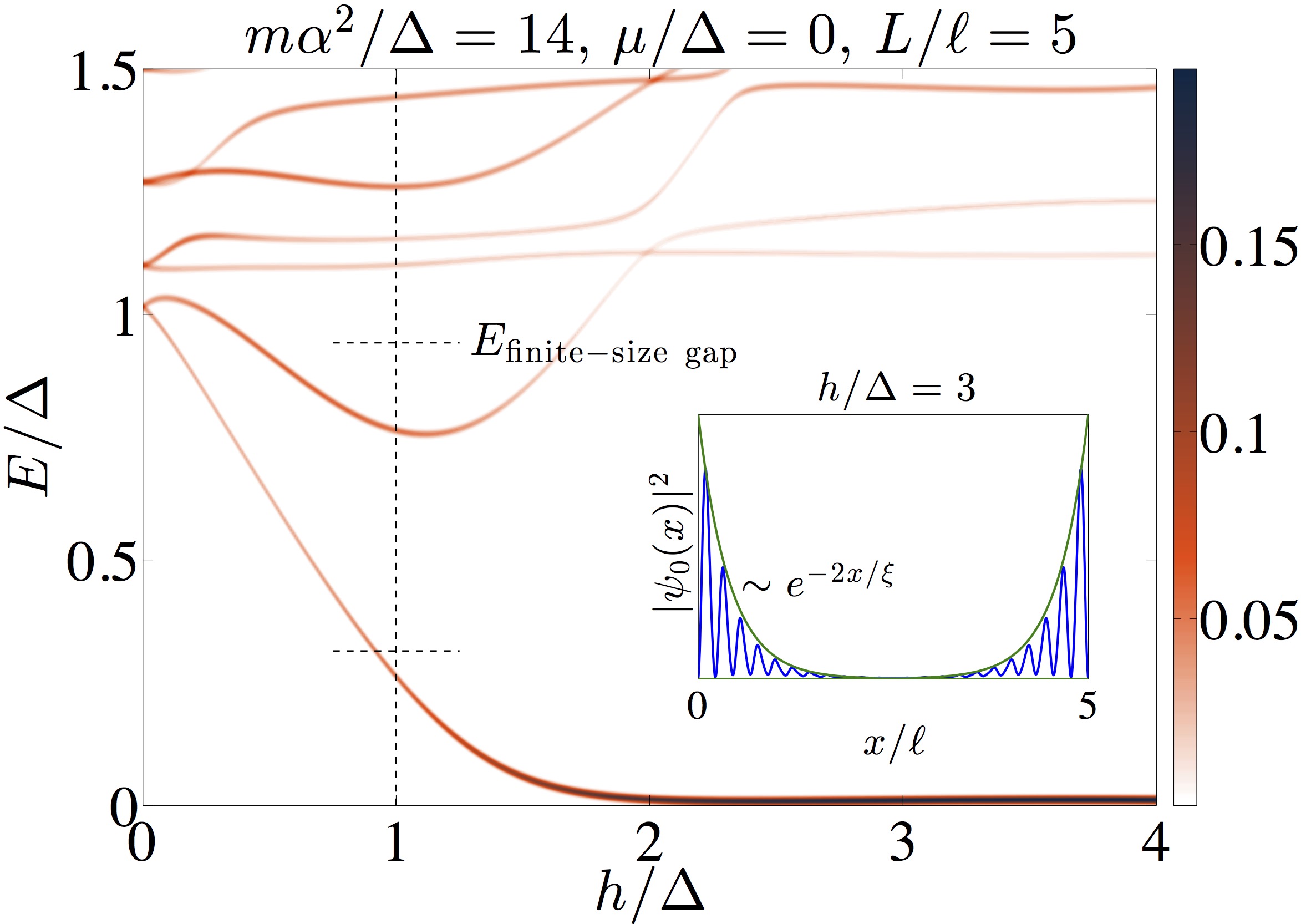}} \hfill
\subfigure{\includegraphics[width=0.315\textwidth]{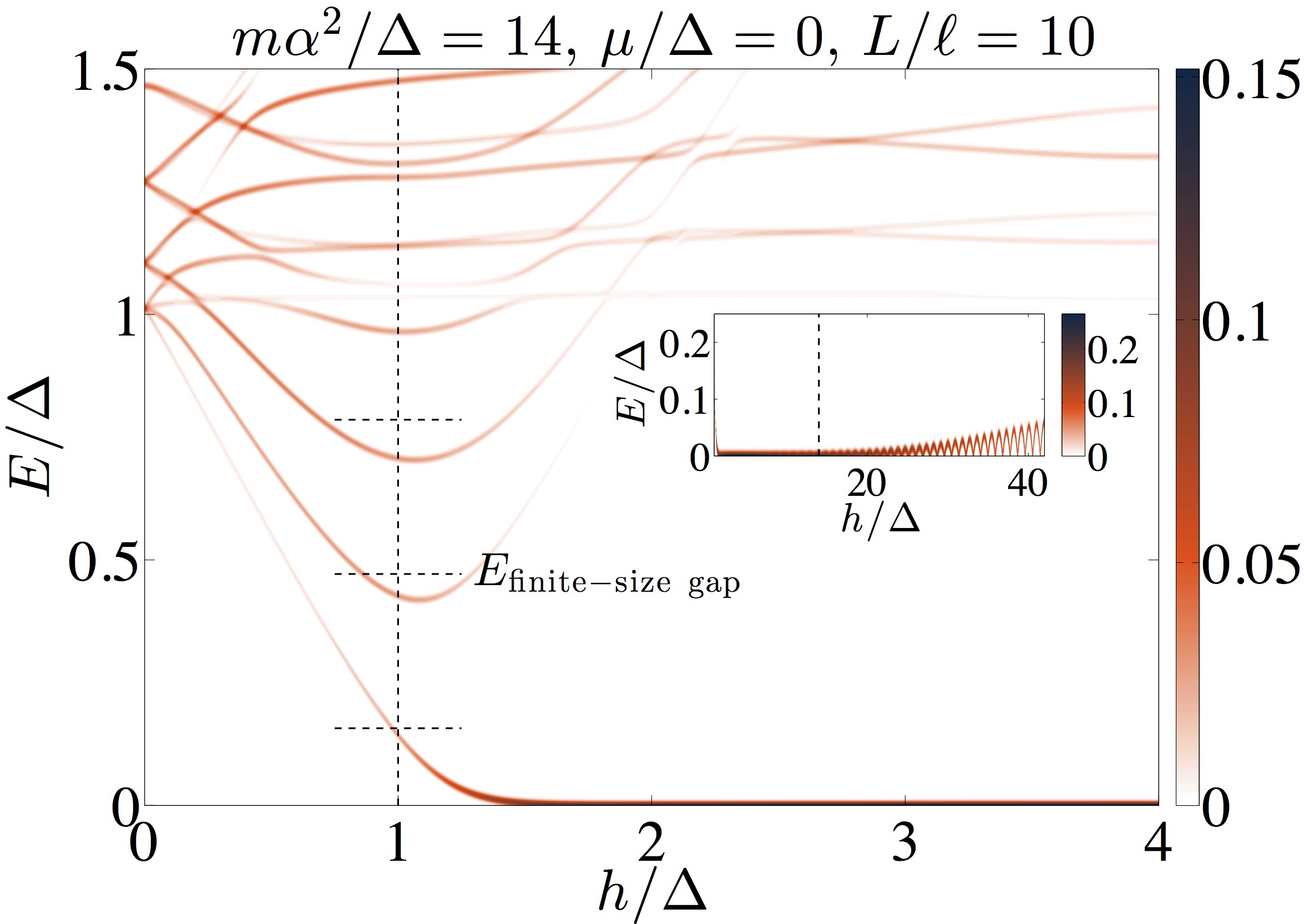}} \hfill
\subfigure{\includegraphics[width=0.315\textwidth]{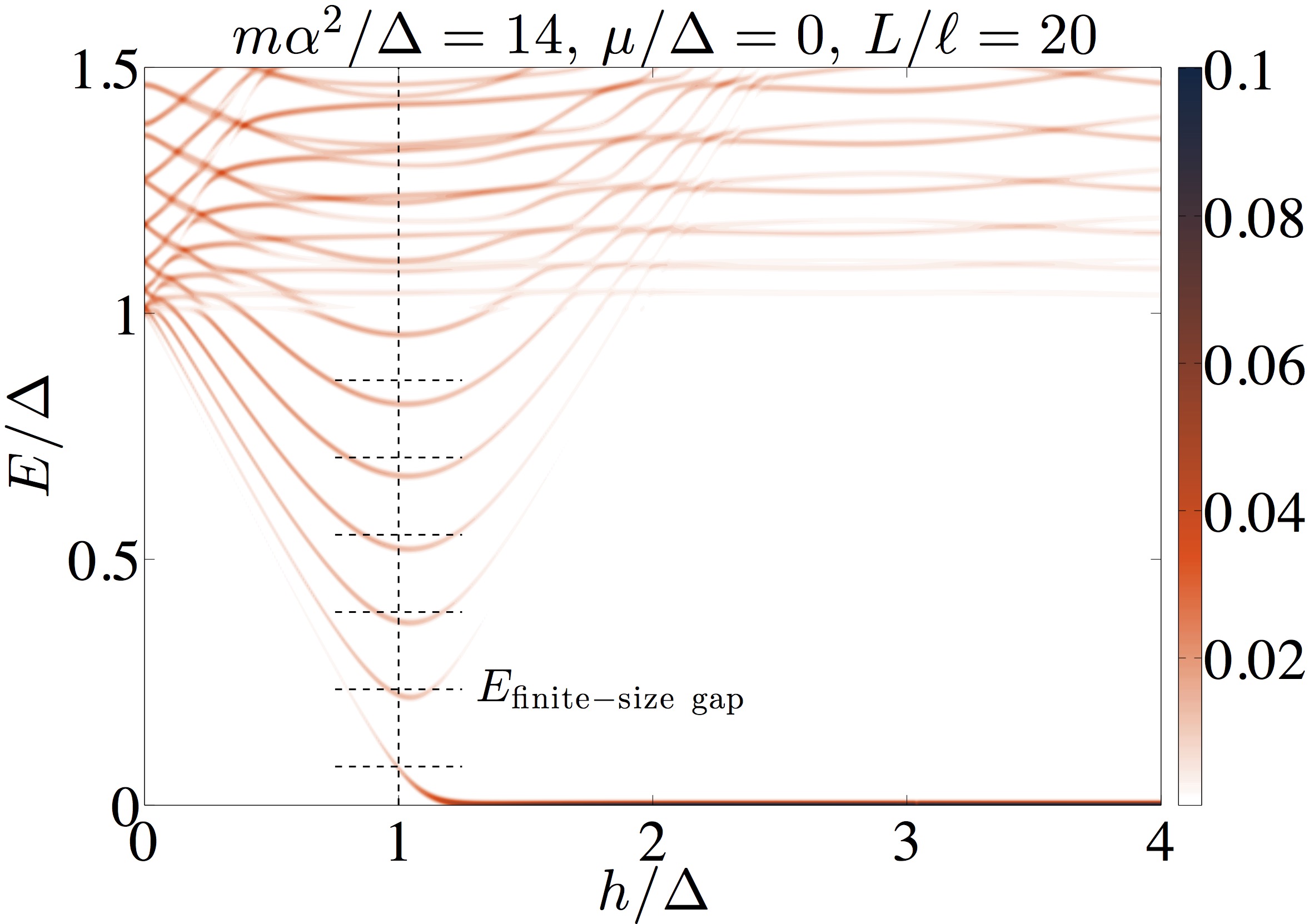}}
}
\centering{
\subfigure{\includegraphics[width=0.315\textwidth]{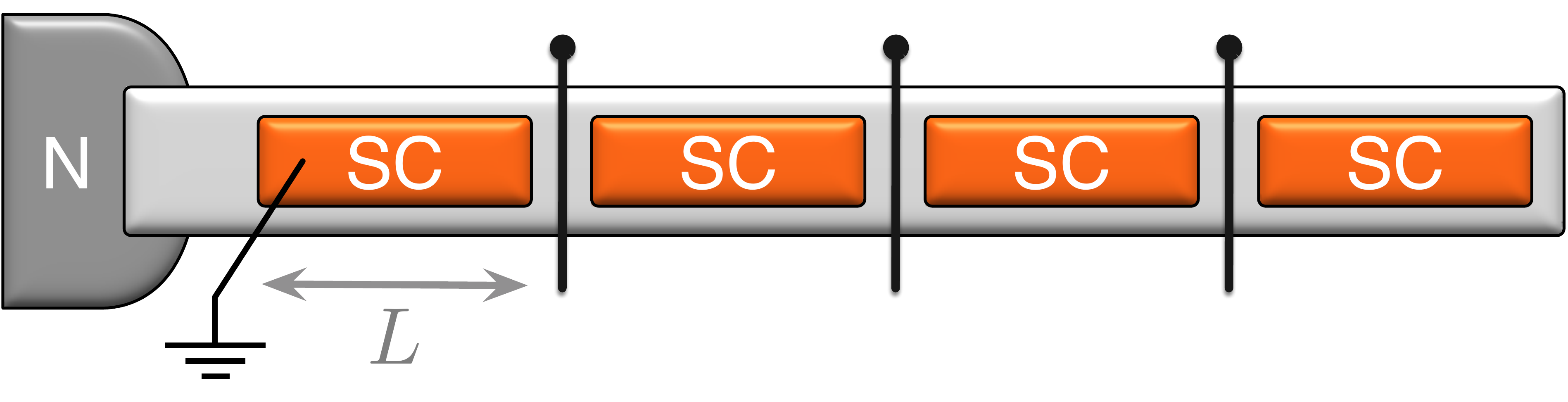}} \hfill
\subfigure{\includegraphics[width=0.315\textwidth]{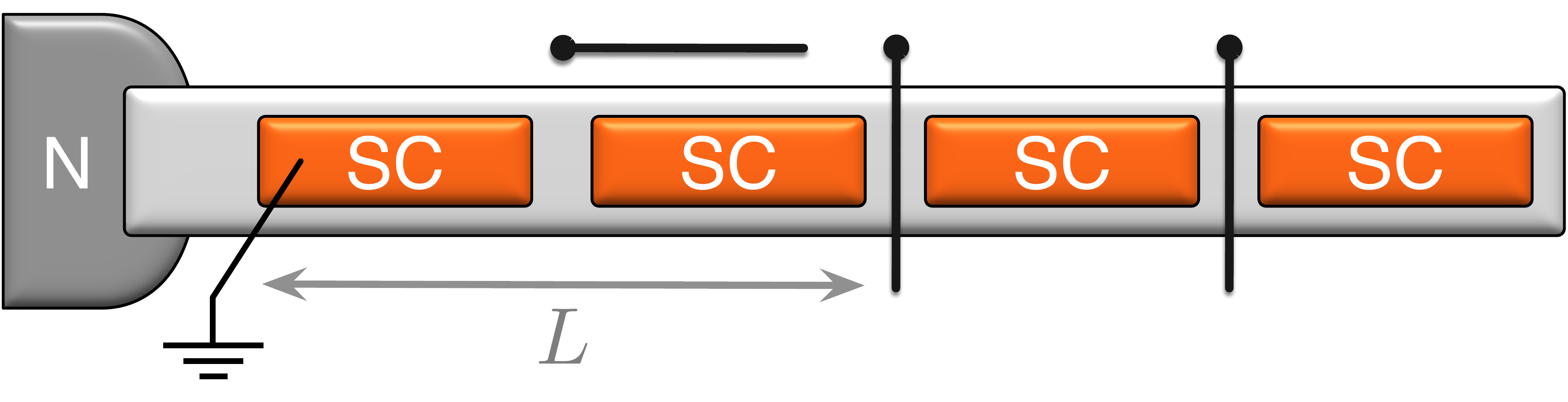}} \hfill
\subfigure{\includegraphics[width=0.315\textwidth]{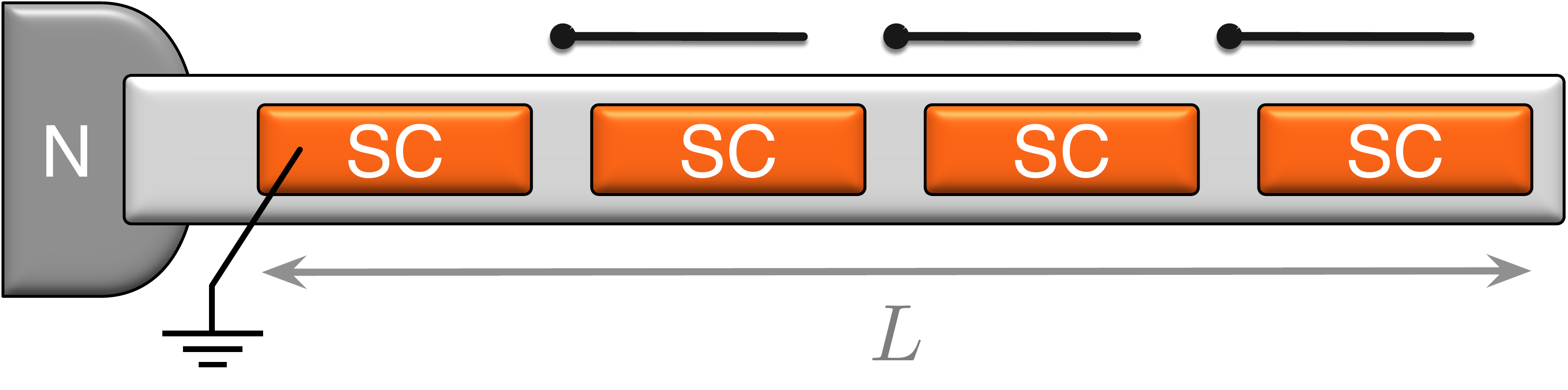}}
}
\caption{Zeeman field scans at strong spin-orbit coupling:  Local density of states (LDOS) [see Eq.~\eqref{eq:LDOS}] versus $h/\Delta$ and $E/\Delta$ with $\mu=0$ in the strong spin-orbit coupling regime, $m\alpha^2/\Delta=14$.  Panels correspond to system sizes $L/\ell=5,10,20$ from left to right, where $\ell \equiv \hbar\alpha/\Delta$ [cf.~Eq.~\eqref{strongSO}].  Data shown represents the usual definition of the LDOS averaged over the leftmost 5\% of the system.  Vertical dashed lines indicate the value $h_c$ of the Zeeman field at the topological phase transition in the thermodynamic limit.  Horizontal lines indicate the levels predicted by Eq.~\eqref{Ediscrete}; for emphasis, we explicitly label the theoretical value $E_{\text{finite-size~gap}} = E_{n = 1}$ given in Eq.~\eqref{Egap1}.  At the smallest system size (left panel), we see an appreciable finite-size bulk gap at $h=h_c$, i.e., $E_{\text{finite-size~gap}}\sim\Delta$, with robust Majorana zero modes still forming deep in the topological phase; in the inset, we show the spatial profile of one of the (nearly) zero-mode wavefunctions at $h/\Delta = 3$ whose envelope follows an exponential with correlation length $\xi\approx0.8\ell$, hence confirming Eq.~\eqref{strongSO}.  For the largest system size (right panel), we are approaching the true quantum critical behavior of the system.  The inset in the middle panel extends for $L/\ell = 10$ the range of Zeeman fields out to $h=3m\alpha^2=42\Delta$ (same data as in the main panel), showing the eventually resolvable splitting and oscillations\cite{SmokingGun} of the zero-bias peak for $h\gtrsim m\alpha^2$ (vertical dashed line).  Bottom panels illustrate the experimental protocol described in Sec.~\ref{sec:FSSprotocol} in which system-size variation can be achieved \emph{in a single device} via gate-tunable valves separating islands of a prescribed length.  Experimentally demonstrating the length dependence shown here would confirm the approach to criticality \emph{and} reveal the spin-orbit strength $\alpha$; recall Eqs.~\eqref{v} and \eqref{Egap1}.}
\label{fig:StrongSO_FieldScans}
\end{figure*}

We propose that the onset of a topological phase transition can be detected by studying conductance spectra in clean, hard-gap wires\cite{Marcus15_NatureMat_14_400, Marcus15_NatureNano_10_232}.  Discrete low-lying sub-gap energy levels for a finite wire should be resolvable through conductance maps as nicely demonstrated, for example, in Ref.~\onlinecite{Mingtang}.  Importantly, these levels exhibit universal characteristics inherited from an infinite system's bona fide phase transition.  We outline several specific experimental protocols designed to probe the imprint of the putative topological phase transition on finite-size wires.  

Inspired largely by the recent developments presented in Refs.~\onlinecite{Mingtang, Gatemon, Gatemon2}, these protocols rely on measurements for systems with either $(i)$ a fixed physical wire length $L$ or $(ii)$ the ability to systematically change $L$ in a single device.  In case $(i)$, we make detailed predictions for the evolution of the finite-system's energy levels near the phase transition---notably their field and density dependence---which may be compared with existing experimental data.  For $(ii)$, we propose varying $L$ using either pinch-off gates between epitaxially grown mesoscopic superconducting islands (see Fig.~\ref{fig:StrongSO_FieldScans} and Refs.~\onlinecite{Gatemon,Gatemon2,Milestones}), or via a more traditional approach \cite{AliceaBraiding} of tuning nearby gates to selectively deplete segments of a wire coupled to a single superconductor.   While more challenging, such experiments can reveal a universal $1/L$ scaling of excited-state energy levels at the phase transition; confirming this behavior would leave little doubt as to the topological nature of the system at higher fields.  The methods we outline further reveal the proximitized wire's spin-orbit coupling strength, an important parameter that has not yet been determined directly in the hybrid structures relevant for Majorana physics.  Overall, we hope that our work helps to inspire further experimental efforts geared towards the unambiguous discovery of a topological phase transition in Majorana nanowires.

\begin{figure*}
\centering{
\subfigure{\includegraphics[width=0.315\textwidth]{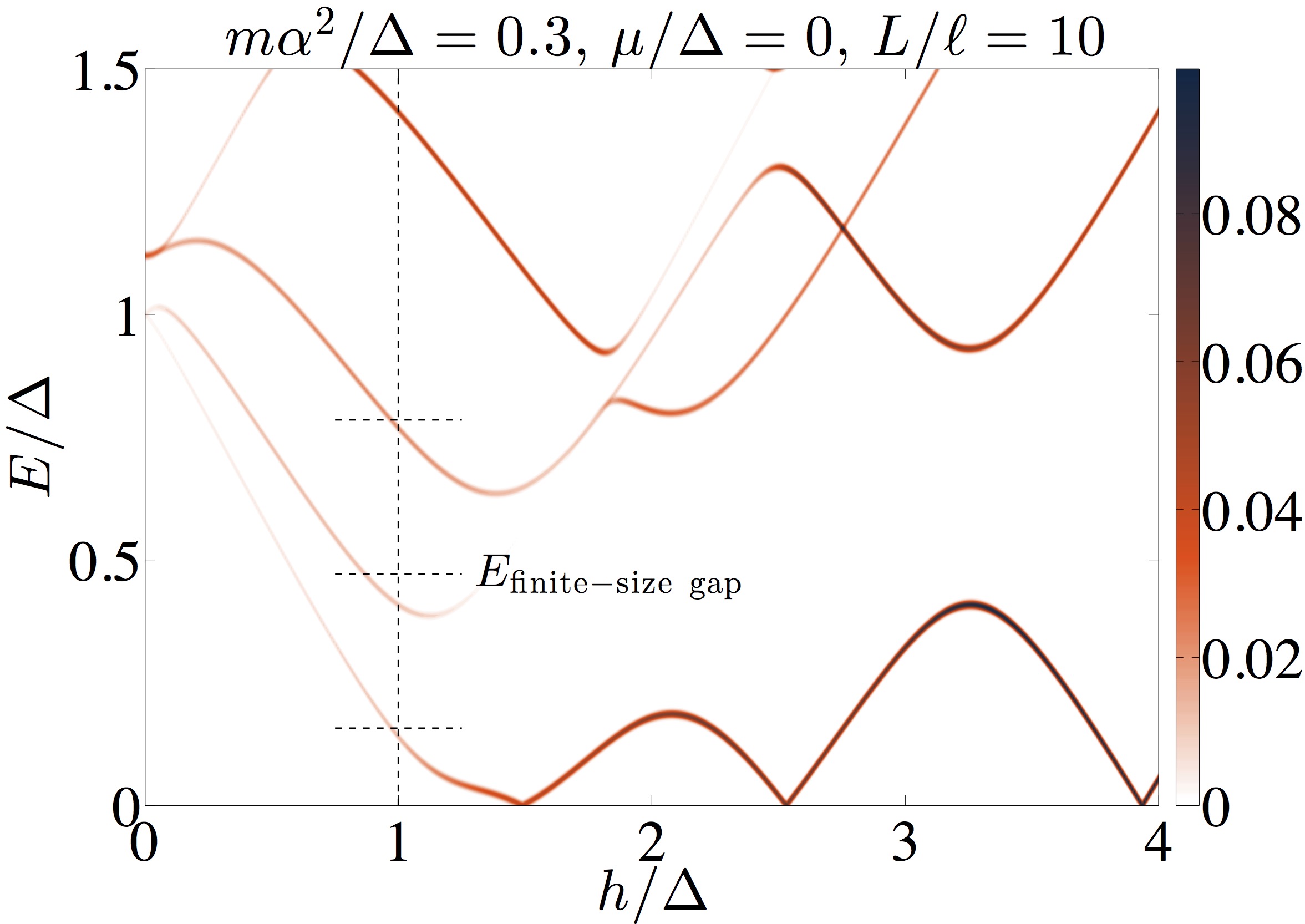}} \hfill
\subfigure{\includegraphics[width=0.315\textwidth]{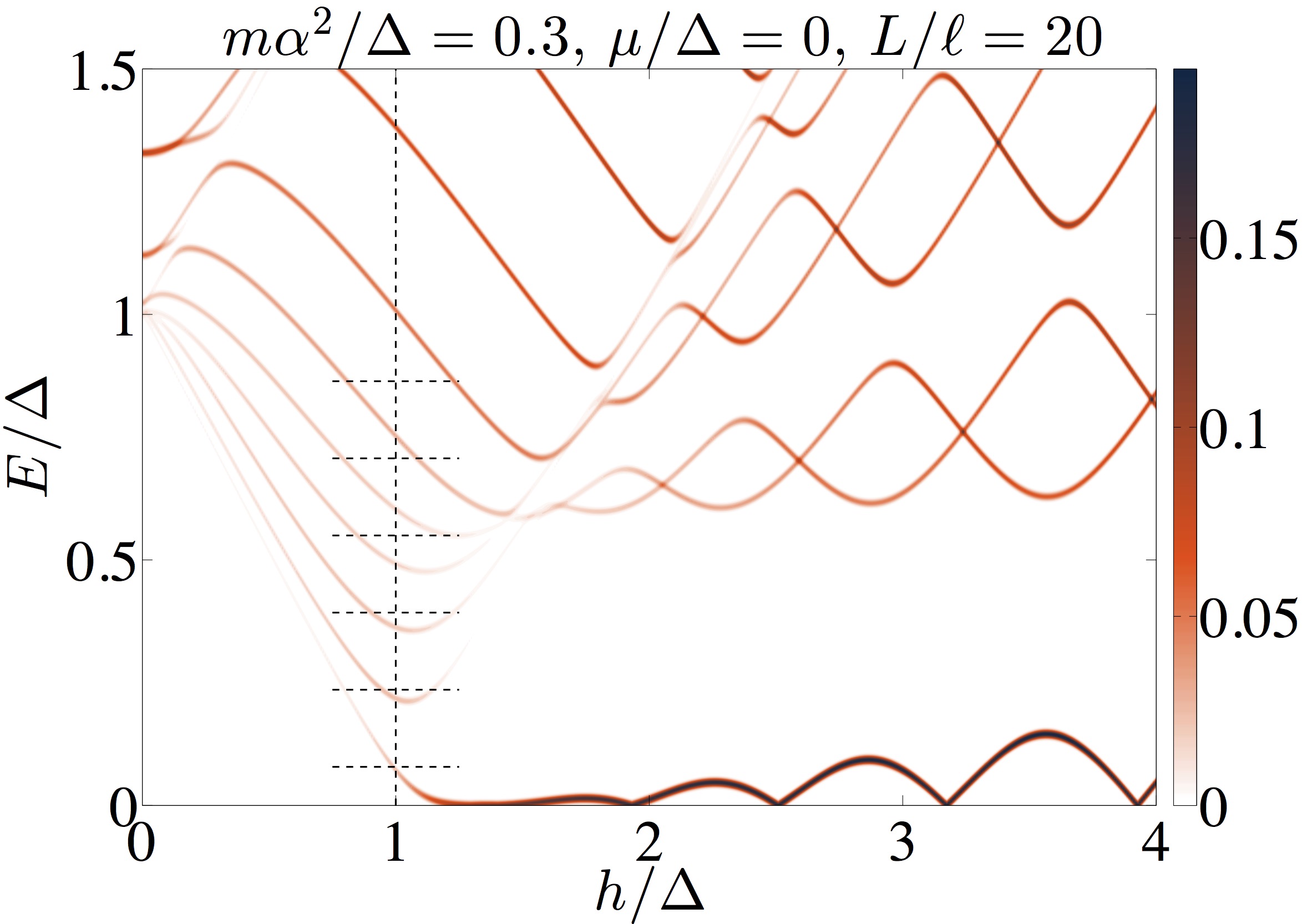}} \hfill
\subfigure{\includegraphics[width=0.315\textwidth]{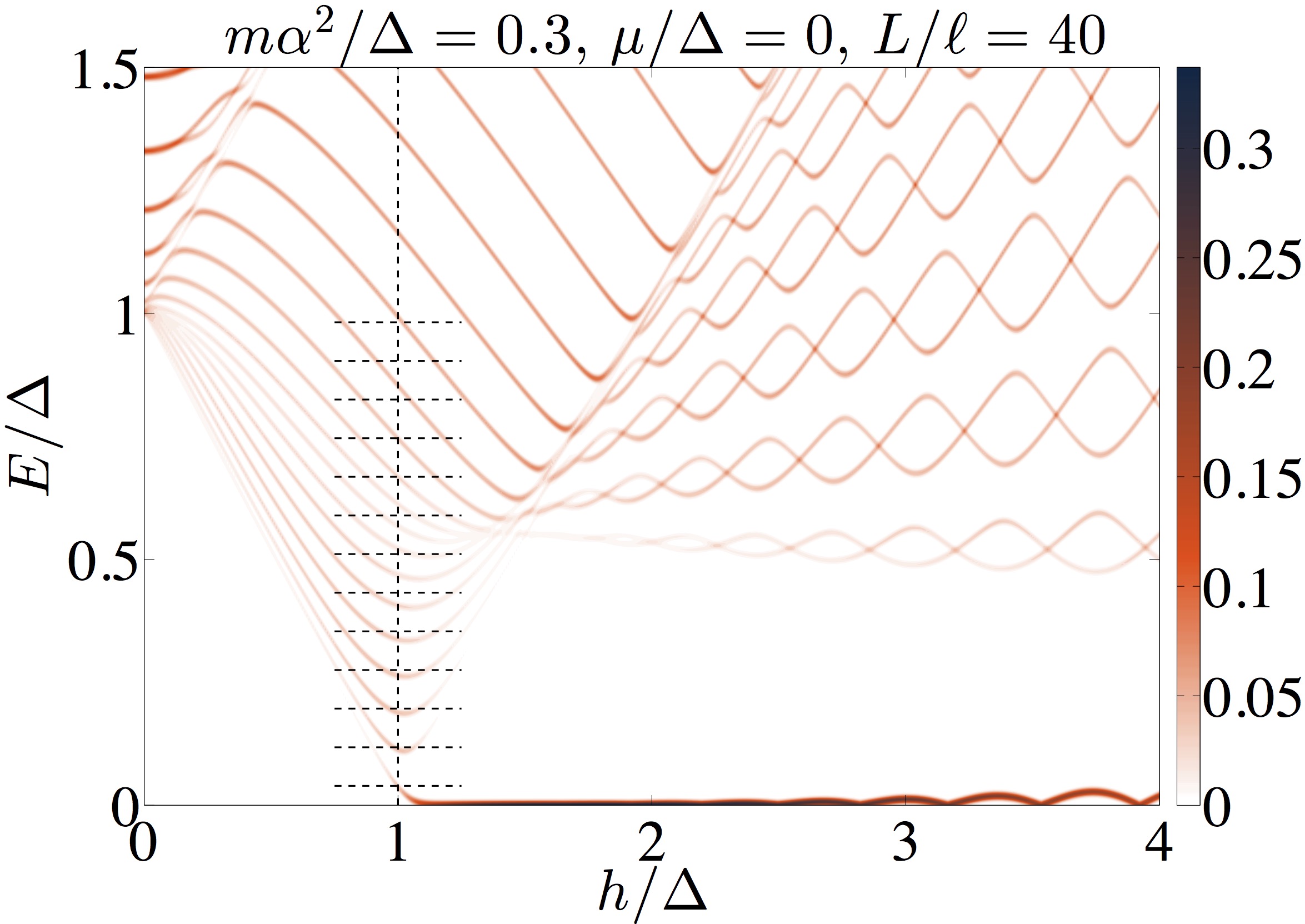}}
}
\caption{Zeeman field scans at weak spin-orbit coupling:  LDOS versus $h/\Delta$ and $E/\Delta$ with $\mu=0$ in the weak spin-orbit coupling regime, $m\alpha^2/\Delta=0.3$.  Panels correspond to system sizes $L/\ell=10,20,40$ from left to right.  Conventions and annotations are the same as in Fig.~\ref{fig:StrongSO_FieldScans}.  This data elucidates the necessity---in the weak spin-orbit regime---of the presence of a sharp topological phase transition if robust zero-modes appear for $h > h_c$.  In the left panel at $L/\ell = 10$, the phase transition is very much a crossover and the zero-bias peak never fully forms.  On the other hand, in the right panel at $L/\ell = 40$, a robust zero-bias peak appears on the topological side of a sharp phase transition.}
\label{fig:WeakSO_FieldScans}
\end{figure*}

\section{Finite-size effects on the topological phase transition} \label{sec:FiniteSize}

For simplicity we consider the minimal single-band model for the superconducting wire.\footnote{Our analysis can be straightforwardly extended to the multi-band case.  Additional bands are, however, expected to be qualitatively unimportant for describing the universal low-energy physics in the vicinity of a topological phase transition.  Roughly speaking, such a transition transpires within a single sub-band, while all others remain fully gapped and are thus largely `spectators'.}  The Hamiltonian reads
\begin{eqnarray}
  H &=& \int_0^L dx\bigg{[}\psi^\dagger \left(-\frac{\hbar^2\partial_x^2}{2m} - \mu -i \hbar \alpha \sigma^y \partial_x + h \sigma^x\right)\psi 
  \nonumber \\
  &+& \Delta (\psi_\uparrow \psi_\downarrow + \mathrm{H.c.})\bigg{]},
  \label{H}
\end{eqnarray}
where $\psi_\alpha$ describes electrons with spin $\alpha$, effective mass $m$, and chemical potential $\mu$; $\sigma^{x,y}$ denote Pauli matrices that act in spin space; $\alpha$ is the spin-orbit strength; $h = \frac{1}{2} g\mu_B B \geq 0$ is the Zeeman energy, with $g$ the wire's effective $g$ factor, $\mu_B$ the Bohr magneton, and $B$ the field strength; and $\Delta$ represents the induced pairing potential.  (Parameters in $H$ should be regarded as effective couplings renormalized by hybridization with the parent superconductor.\cite{Disorder4})  Consider first an infinite system.  Within this model the phase transition occurs when $h = h_c \equiv \sqrt{\Delta^2 + \mu^2}$.\cite{1DwiresLutchyn,1DwiresOreg}  This condition can be satisfied by tuning either the field or chemical potential, and our analysis of the level structure at and near the topological quantum critical point holds independent of which control parameter is varied.  In the case of chemical-potential tuning, the analysis applies equally well to either of the two critical points that border the topological phase on its low- and high-density sides (see Fig.~\ref{fig:muScans}).  By diagonalizing Eq.~\eqref{H} one finds that at criticality the low-energy excitation spectrum is
\begin{equation}
  E_k = \hbar v |k|
  \label{Ek}
\end{equation}
with $k$ the momentum and
\begin{equation}
  v = \alpha \frac{\Delta}{\sqrt{\Delta^2 + \mu^2}},
  \label{v}
\end{equation}
a velocity bounded by the spin-orbit strength, i.e., $v \leq \alpha$.  

Suppose now that the superconducting wire has a finite length $L$.  In this case it is natural to expect the continuous energy spectrum in Eq.~\eqref{Ek} to become discrete due to finite-size momentum quantization.  Appendix \ref{TransitionAppendix} shows that the allowed momenta indeed become $k_n = \frac{\pi}{L}(n+\frac{1}{2})$ for integer $n$.  Inserting $k_n$ into Eq.~\eqref{Ek} yields quantized energy levels
\begin{equation}
  E_n = \frac{\pi \hbar v}{L}\left(n+\frac{1}{2}\right),~~~~~n = 0, 1,2,\cdots.
  \label{Ediscrete}
\end{equation}
The lowest $n = 0$ bulk mode evolves into localized Majorana end-states in the adjacent topological phase (see Fig.~\ref{fig:StrongSO_FieldScans}).  We are interested primarily in the next excited state, whose energy at criticality is given by
\begin{equation}
  E_{\text{finite-size~gap}} \equiv E_{n = 1} = \frac{3\pi \hbar v}{2L}.
  \label{Egap1}
\end{equation}
This level generally corresponds to the lowest-lying mode that is a bulk state on both sides of the transition.  In the literature, sometimes Ôgap closureÕ at the topological phase transition refers simply to continuous formation of a zero-bias peak upon increasing field, i.e., $E_{n = 0}\to 0$. For quantifying how well a finite-size system approximates true quantum critical behavior, however, it is very useful to consider the behavior of this next excited state $E_{n = 1}$---which reveals not only the closing but also the crucial reopening of the bulk gap (precisely the $E_{n=1}$ level) upon entering the topological regime\footnote{In the thermodynamic limit, there is a continuum of states above the bulk gap on both the trivial and topological sides of the phase transition (and at the transition itself).  The bulk gap always corresponds to the level $E_{n=1}$ in the topological phase, as well as at the transition where it vanishes as $L\to\infty$.}.

Now, to get a sense of scales, consider a wire with length $L \sim 1~\mu$m comparable to the largest superconducting segments studied in previous experiments \cite{mourik12,das12,deng12,finck12,Churchill}.  We further assume for now that the chemical potential satisfies $\mu \lesssim \Delta$ so that the topological regime appears in fairly low fields.  The Rashba spin-orbit strengths have not been measured in proximitized wires---for which the adjacent superconductor can contribute appreciably---though based on previous measurements for bare nanowires \cite{InSbSOC,Nadj-Perge12_PRL_108_166801,Nadj-Perge10_PhDThesis,Petta11_PRL_107_176811, Loss07_PRL_98_266801,Nadj-Perge10_Nature_468_1084} we expect that $\alpha \sim 10^4-10^5~\mathrm{m/s}\sim0.07-0.7~\mathrm{eV}\,\mathrm{\AA}/\hbar$ is reasonable.  With $\alpha$ near the middle of this range, the residual bulk gap at the `phase transition' is then
\begin{equation}
  E_{\text{finite-size~gap}} \sim 1~{\rm K},
  \label{Egap2}
\end{equation}
indeed comparable to the induced pairing energies deduced experimentally \cite{mourik12,das12,deng12,finck12,Churchill}.  

If finite-size effects obscure the phase transition to the extent suggested by Eq.~\eqref{Egap2}, can well-formed Majorana modes still appear?  The answer depends sensitively on the spin-orbit strength, quantified by the energy scale $m\alpha^2$.  

\subsection{Weak spin-orbit coupling: $m\alpha^2 \ll \Delta$}

Weak spin-orbit coupling corresponds to $m\alpha^2 \ll \Delta$.  In this regime the topological phase's correlation length, which determines the spatial decay of Majorana-zero-mode wavefunctions, is approximately
\begin{equation}
  \xi \sim \left(\frac{h}{m\alpha^2}\right)\frac{\hbar \alpha}{\Delta} ~~~~~~\text{(weak spin-orbit)}.
  \label{weakSO}
\end{equation}
This estimate---as well as Eq.~\eqref{strongSO} below---simply follows from the ratio of the Fermi velocity to the bulk gap and holds provided the system is not too close to the transition (where $\xi$ formally diverges).  Importantly, since the topological phase requires $h>\Delta$, the prefactor in parenthesis above is large in the weak spin-orbit regime.  Using Eqs.~\eqref{v} and \eqref{Egap1}, we then see that having $E_{\text{finite-size~gap}}$ comparable to $\Delta$ implies that $\xi \gtrsim L$.  Robust Majorana modes are then generically absent in this scenario since their spatial extent would exceed the system size.  Equivalently, observing a Majorana-induced zero-bias peak in a weakly spin-orbit-coupled wire would necessitate a rather sharp topological phase transition detectable by some means.

We confirm this behavior numerically in Fig.~\ref{fig:WeakSO_FieldScans} (see Appendix~\ref{NumericalDetails} for numerical details) where we plot the local density of states (LDOS) averaged over the leftmost 5\% of the wire versus Zeeman strength, $h/\Delta$, and energy, $E/\Delta$.  In these simulations we choose parameters $\mu=0$ and $m\alpha^2/\Delta=0.3$---deep in the weak spin-orbit regime---and use as a reference length $\ell\equiv\hbar\alpha/\Delta$.  The three panels correspond to system sizes $L/\ell = 10, 20, 40$ from left to right.   
For the shortest system, $L/\ell = 10$, we indeed have $E_{\text{finite-size~gap}}\sim\Delta$ with strongly split zero modes for $h > h_c = \Delta$.  It is not until the wire is sufficiently long, e.g., at $L/\ell = 40$, that a robust zero-bias peak forms after crossing a nearly genuine thermodynamic topological phase transition with its associated pile-up of energy levels at the critical point.

\subsection{Strong spin-orbit coupling: $m\alpha^2 \gg \Delta$}

Qualitatively different physics emerges at strong spin-orbit coupling, i.e., $m\alpha^2 \gg \Delta$.  Here the topological phase enjoys a broad field range, extending from $h \sim \Delta$ to $h\sim m \alpha^2$, where the modes gapped by Cooper pairing carry nearly antiparallel spins---thereby maximizing the bulk gap.  
Within this field window a parametrically shorter correlation length arises,
\begin{equation}
  \xi \sim \frac{\hbar \alpha}{\Delta} ~~~~~~\text{(strong spin-orbit)},
  \label{strongSO}
\end{equation}
which again holds not too close to the transition.  [For $h \gtrsim m \alpha^2$ the Zeeman field begins to overwhelm the spin-orbit energy, yielding a smaller bulk gap and correspondingly larger correlation length that eventually recovers Eq.~\eqref{weakSO}---see inset of Fig.~\ref{fig:StrongSO_FieldScans}, middle panel.]  Systems for which $E_{\text{finite-size~gap}}$ approaches $\Delta$ can consequently still support localized Majorana modes with $\xi \ll L$ over an extended field interval.  For a quantitative estimate, the splitting that arises from overlap of Majoranas bound to opposite ends of the wire is (modulo oscillatory corrections)\cite{SmokingGun}
\begin{equation}
E_{\rm splitting} \sim \Delta e^{-L/\xi}.
\label{Esplitting}
\end{equation}
Taking $E_{\text{finite-size~gap}} = \Delta$ and $\mu = 0$ then yields $E_{\rm splitting} \sim \Delta e^{-3\pi/2} \approx 0.01\Delta$---well below the bulk gap.  \emph{Thus strong spin-orbit coupling allows rather robust localized Majorana modes to appear even in `small' systems where the topological phase transition is severely obliterated into a crossover.}

To drive home this key point in more detail, we now present numerical simulations of Eq.~\eqref{H} in the strong spin-orbit regime.  We again set $\mu = 0$ but now assume strong spin-orbit coupling with $m\alpha^2/\Delta=14$.  Figure~\ref{fig:StrongSO_FieldScans} displays the end-of-wire LDOS for system sizes $L/\ell = 5, 10, 20$ from left to right.   For wurtzite InAs wires with \cite{WurtziteMass} $m = 0.05 m_e$ ($m_e$ denotes the bare electron mass) and $\Delta = 1.5~\mathrm{K}=130~\mu\mathrm{eV}$ the parameters specified above correspond approximately to systems of length $L=2,4,8~\mu\mathrm{m}$ with Rashba coupling \footnote{While \emph{bulk} wurtzite and zincblende band structures can be quite different, we expect Eq.~\eqref{H} to describe reasonably well the highest partially occupied bands in quantum wires of either type.  The form of the Hamiltonian---which one should view as a gradient expansion for low-lying wire states near the zone center---is indeed much more strongly constrained in 1D compared to in higher dimensions.} $\alpha=8\cdot10^4~\mathrm{m}/\mathrm{s}=0.5~\mathrm{eV}\,\mathrm{\AA}/\hbar$.  We believe these are reasonable numbers for Majorana experiments \footnote{For reference, the typically quoted parameter values for the original Delft experiment \cite{mourik12} using InSb wires ($m=0.015m_e$, $\alpha=0.2~\mathrm{eV}\,\mathrm{\AA}/\hbar=3\cdot10^{4}~\mathrm{m}/\mathrm{s}$, and $\Delta=250~\mu\mathrm{eV}=2.9~\mathrm{K}$) result in $m\alpha^2/\Delta=0.3$---corresponding to the weak spin-orbit regime, cf.~Fig.~\ref{fig:WeakSO_FieldScans}.  With these parameters at, say, $h = 2\Delta$ the correlation length predicted by Eq.~\eqref{weakSO} is $\xi \approx 0.5~\mu$m.  A device of length $L = 1~\mu\mathrm{m}$ translates to $L/\ell = 12.5$, which lies near the left panel of Fig.~\ref{fig:WeakSO_FieldScans}.  There is, however, considerable experimental uncertainty in the value of the Rashba coupling, and since the spin-orbit energy $\sim\alpha^2$ such uncertainty can play a large role.} based on the technology developed in Refs.~\onlinecite{Marcus15_NatureMat_14_400, Marcus15_NatureNano_10_232}.  

In each panel of Fig.~\ref{fig:StrongSO_FieldScans}, the vertical dashed line indicates the location of the phase transition for an infinite system, while the horizontal lines denote the levels predicted by Eq.~\eqref{Ediscrete}.  [Deviations between numerics and our analytical prediction arise from higher-order terms not included in Eq.~\eqref{Ek}, which become progressively less important as the wire length increases; see Fig.~\ref{ScalingFig}.]  For the shortest system size in Fig.~\ref{fig:StrongSO_FieldScans} (left panel, $L/\ell=5$), we see precisely the interesting scenario described above where the phase transition is completely smoothed into a crossover, i.e., $E_{\text{finite-size~gap}}\sim\Delta$, yet robust Majorana modes develop in the topological regime.  Similar plots appear in, for example, Refs.~\onlinecite{DimensionalCrossover,Loss13_PRB_87_024515} where the evolution of the Majorana mode in finite-size systems---rather than excited states at the transition---was studied.

It is worth stressing that while a mere two sub-gap states appear in the crossover region, the Majorana modes remain well localized.  This feature is demonstrated explicitly in the inset of the left panel of Fig.~\ref{fig:StrongSO_FieldScans}, where we plot the wavefunction amplitude $|\psi_0(x)|^2=\sum_{s=\uparrow,\downarrow}\left[|u_0^s(x)|^2 + |v_0^s(x)|^2\right]$ at $h/\Delta=3$ for the lowest-lying eigenstate of Eq.~\eqref{H}; here $u_0^s$ and $v_0^s$ respectively denote components in the particle and hole sectors (see also Appendix~\ref{NumericalDetails}).  Indeed, the probability weight exponentially localizes to the edges and fits very well to a form $|\psi_0(x)|^2\sim e^{-2x/\xi} + e^{-2(L-x)/\xi}$, with correlation length $\xi\approx0.8\ell = 0.8\,\hbar\alpha/\Delta$ that agrees well  with the estimate from Eq.~\eqref{strongSO}.  Furthermore, we have verified that the resolved zero-mode splitting at $h/\Delta=3$ in the left panel of Fig.~\ref{fig:StrongSO_FieldScans} is fully consistent with this correlation length inserted into Eq.~\eqref{Esplitting}.

As we approach the thermodynamic limit by increasing the system size to $L/\ell=10,20$ (middle and right panels of Fig.~\ref{fig:StrongSO_FieldScans}), the finite-size energy levels at the crossover decrease as expected from Eqs.~\eqref{Ediscrete} and \eqref{Egap1}, and the approach to true criticality becomes evident.  At $L/\ell=20$ we obtain very good agreement with the theoretical prediction of Eq.~\eqref{Ediscrete}.  

\begin{figure*}
\centering{
\subfigure{\includegraphics[width=0.315\textwidth]{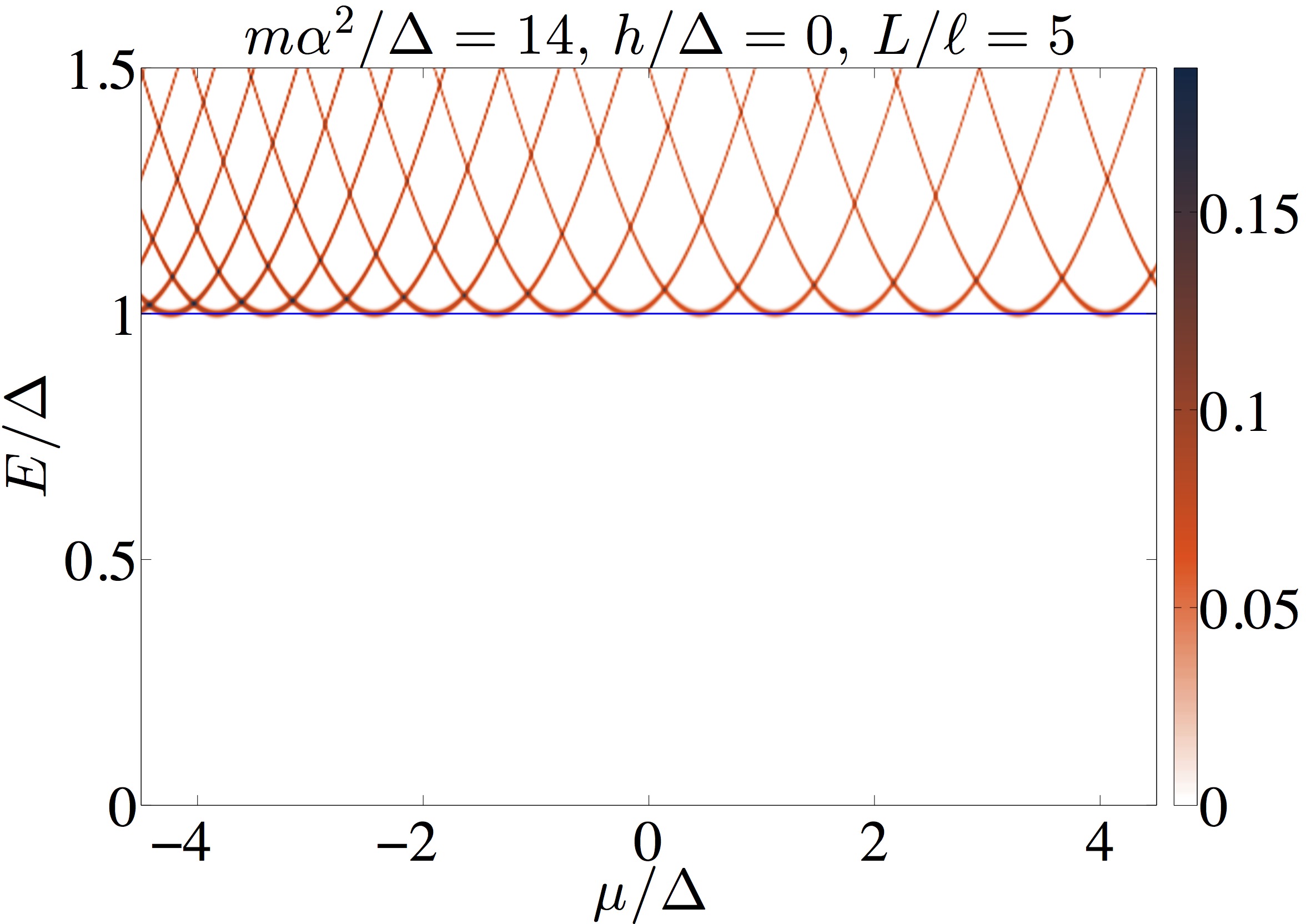}} \hfill
\subfigure{\includegraphics[width=0.315\textwidth]{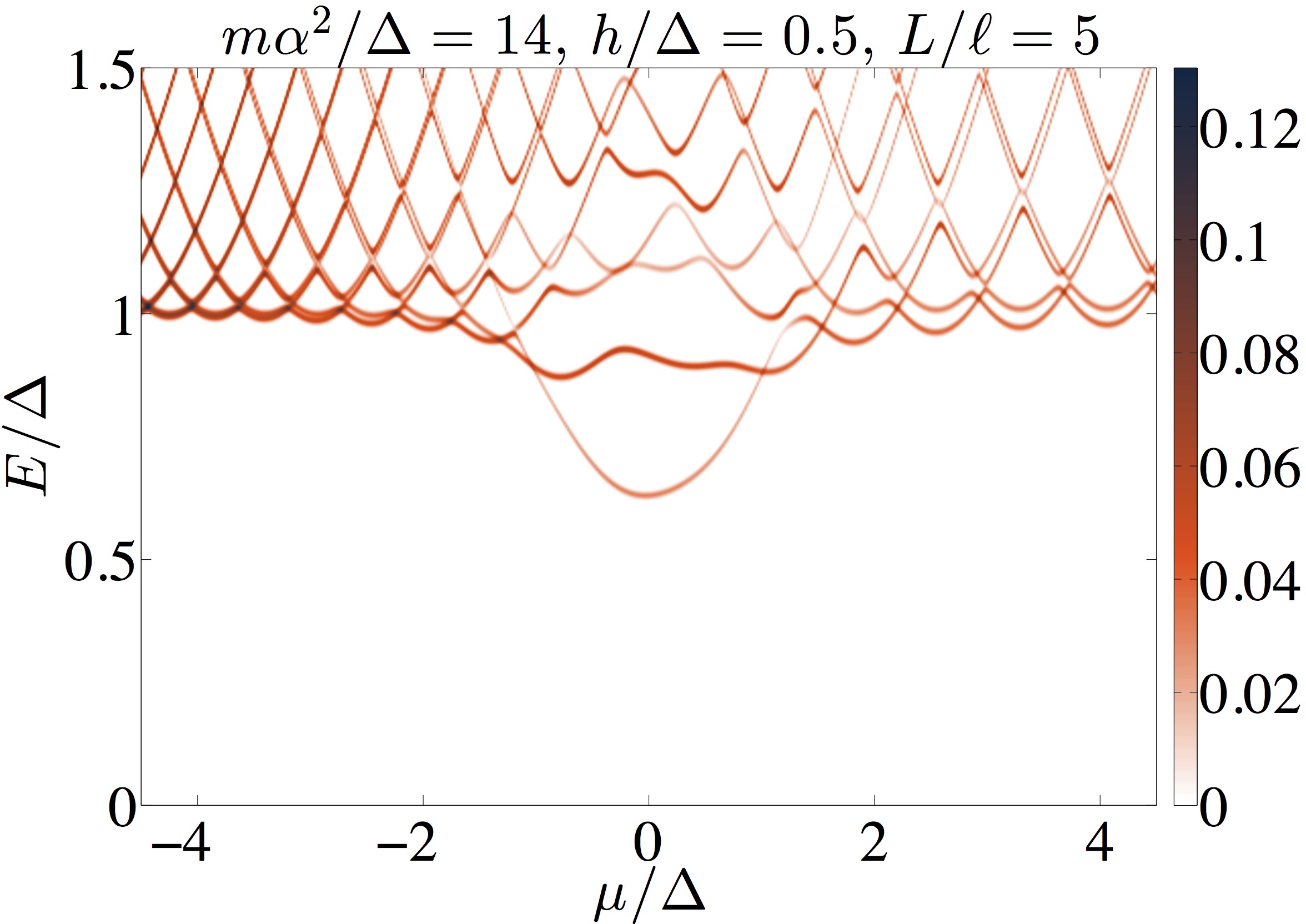}} \hfill
\subfigure{\includegraphics[width=0.315\textwidth]{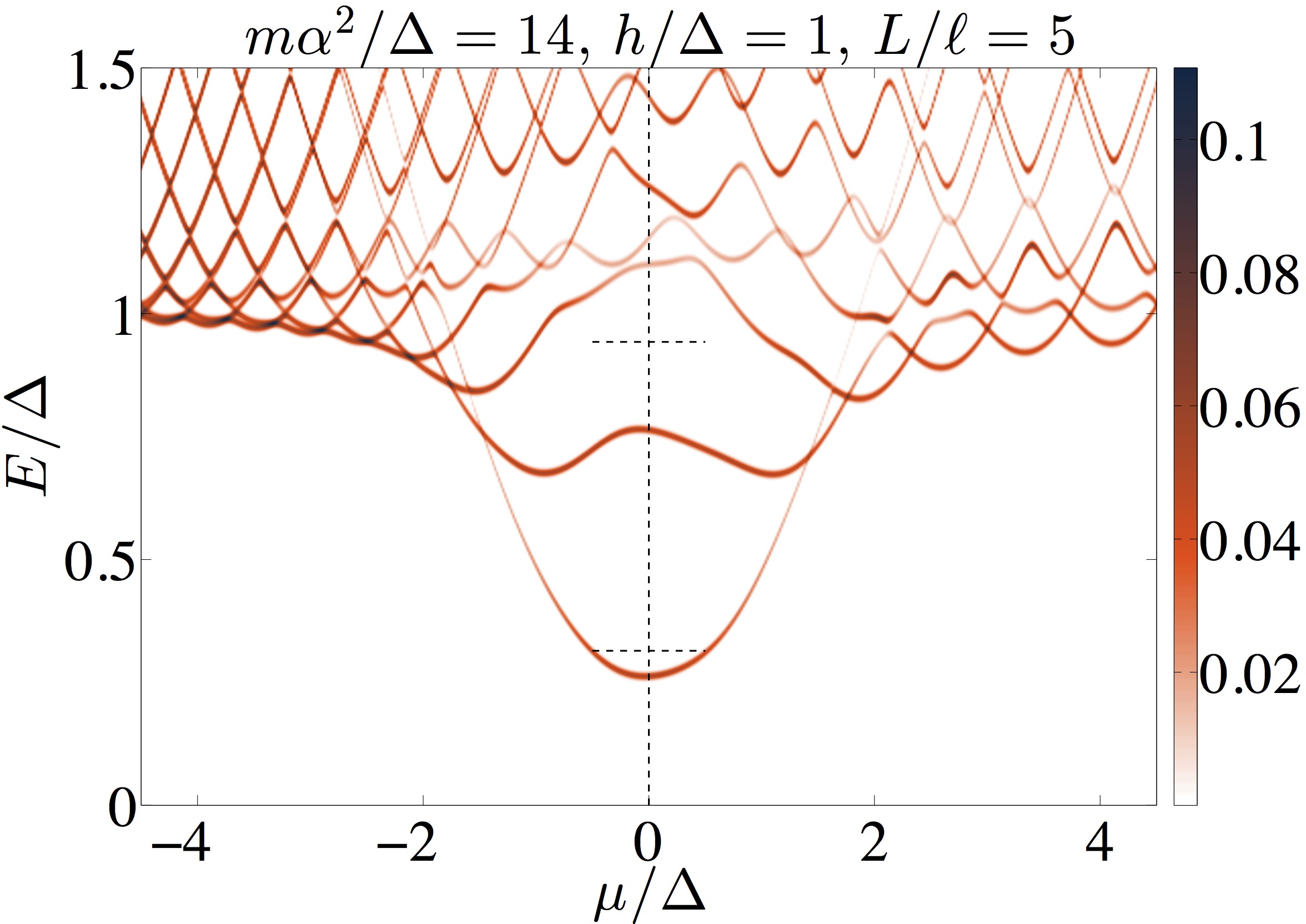}}
}
\centering{
\subfigure{\includegraphics[width=0.315\textwidth]{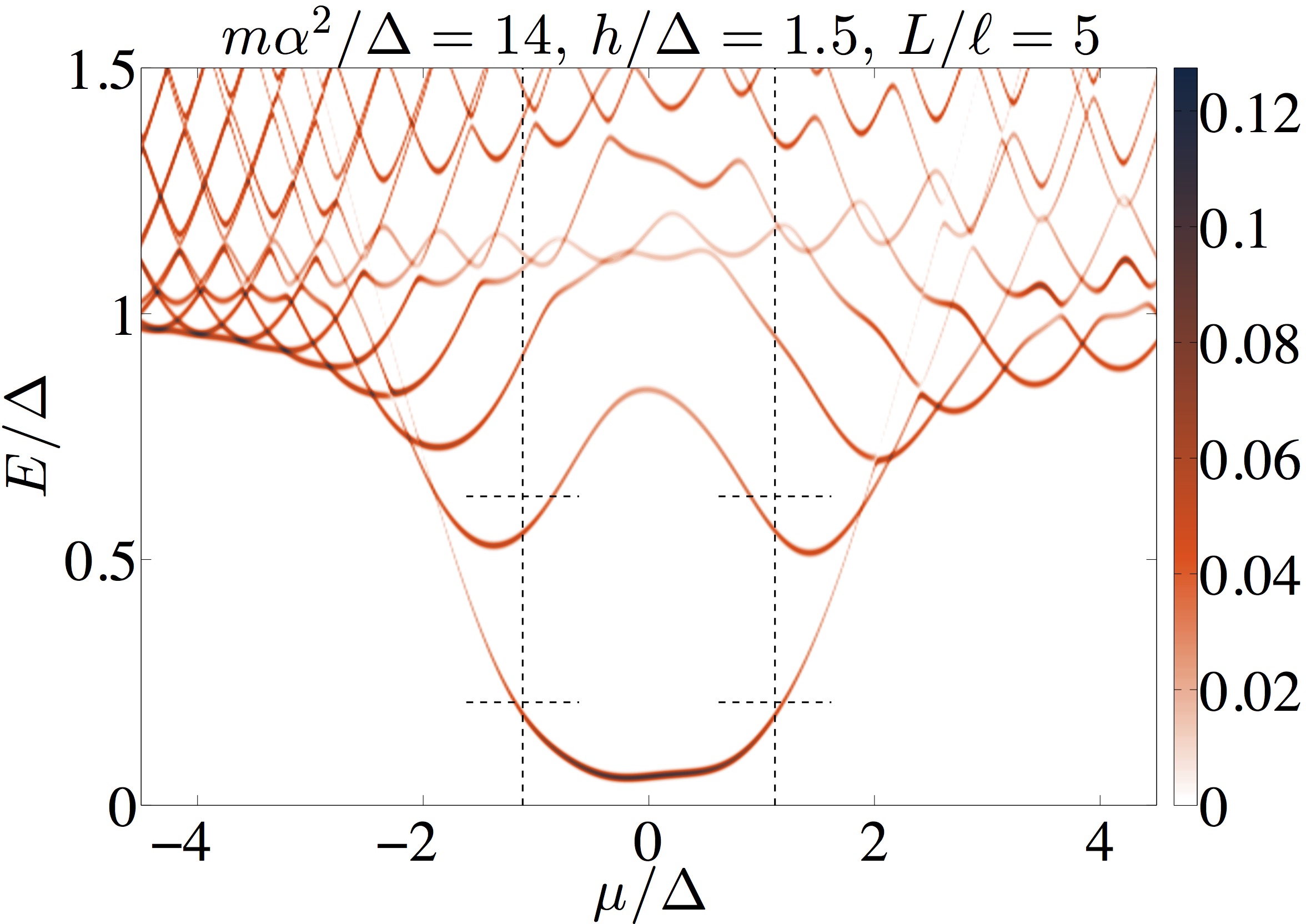}} \hfill
\subfigure{\includegraphics[width=0.315\textwidth]{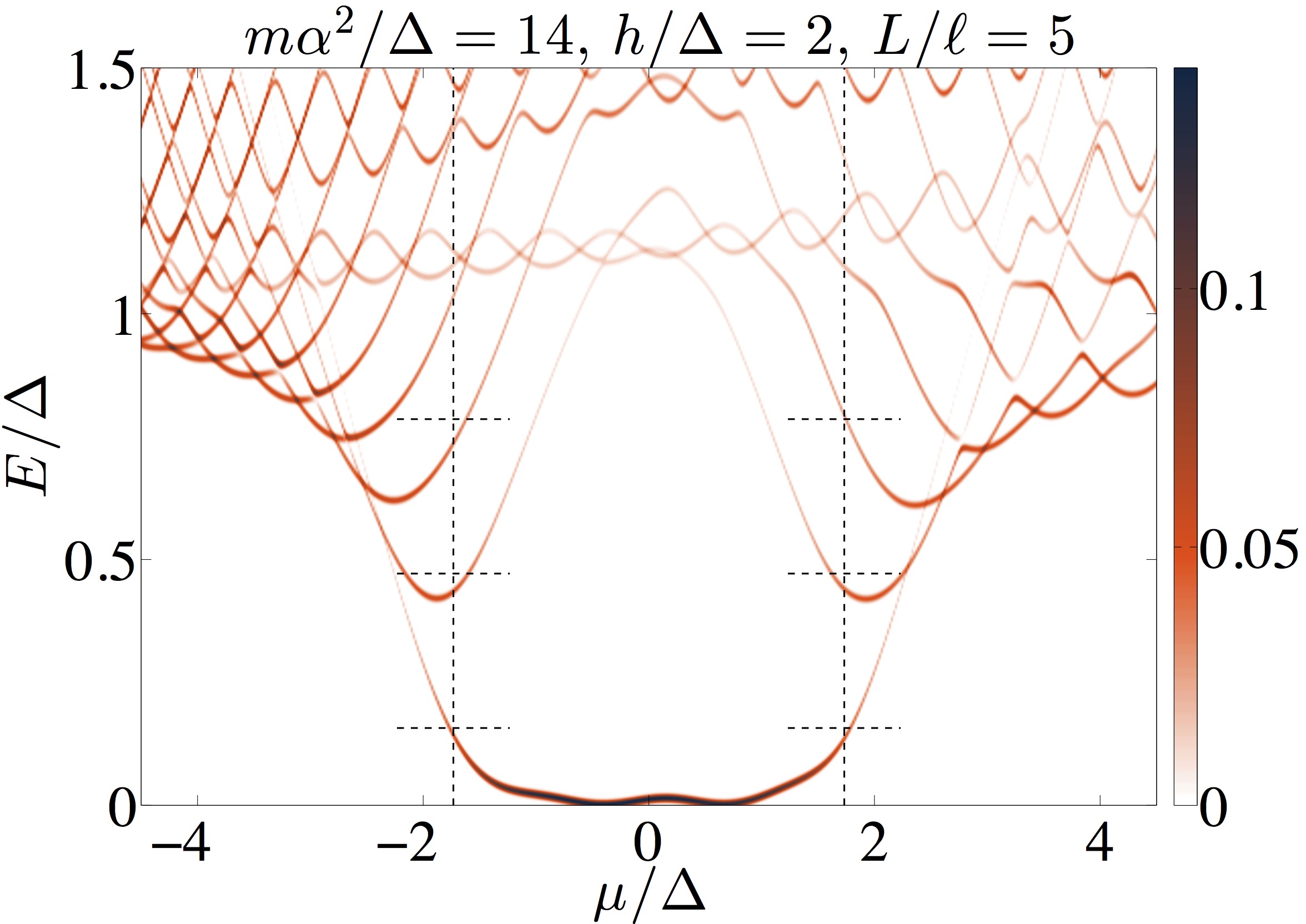}} \hfill
\subfigure{\includegraphics[width=0.315\textwidth]{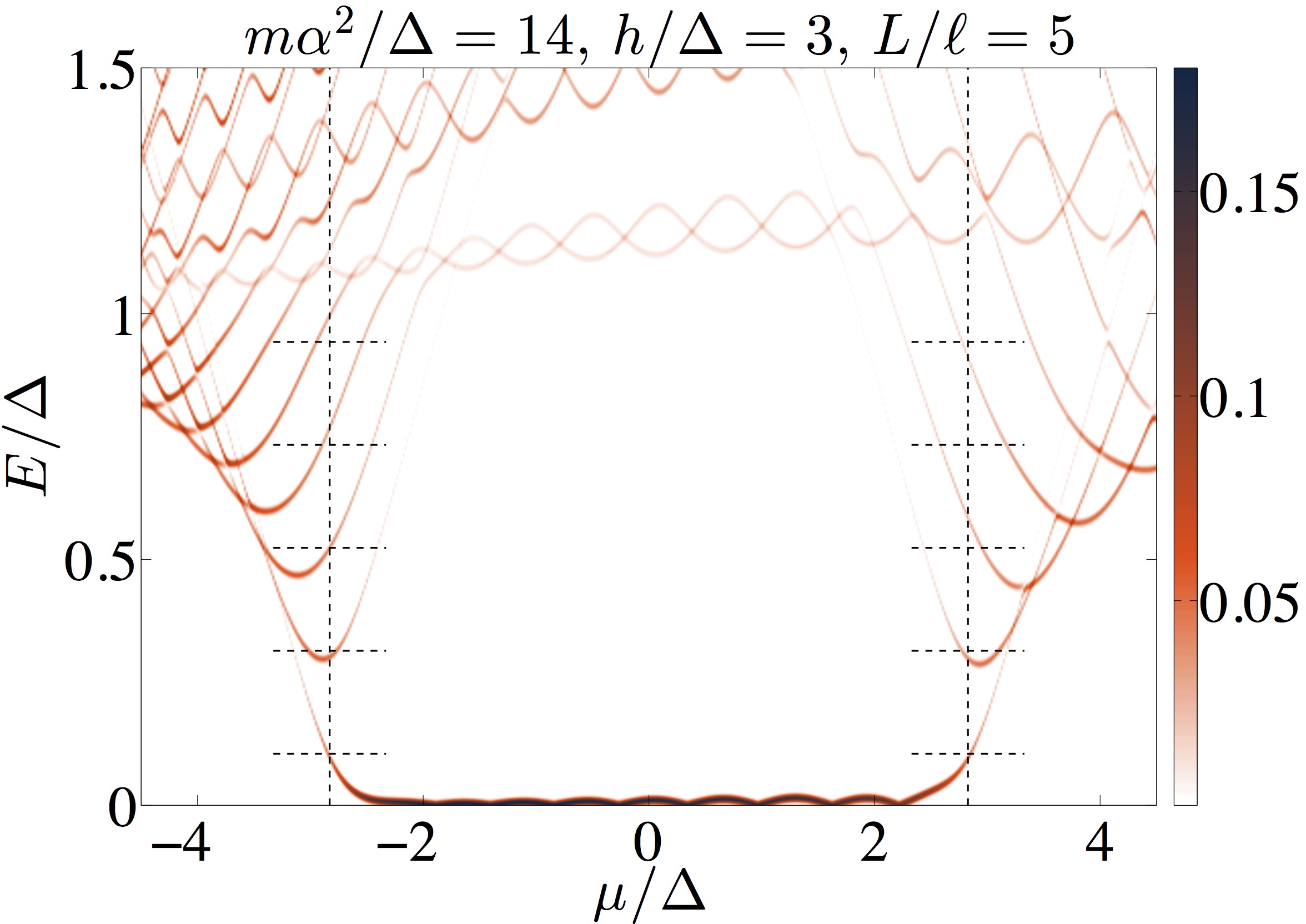}}
}
\caption{Chemical potential scans at strong spin-orbit coupling:  LDOS versus $\mu/\Delta$ and $E/\Delta$ on a system of length $L/\ell=5$ in the strong spin-orbit coupling regime, $m\alpha^2/\Delta=14$.  Different panels correspond to different values of the Zeeman energy $h/\Delta$ as indicated above each plot.  The induced gap $\Delta$ is indicated with a blue horizontal line in the first panel.  The outermost vertical dashed lines, if they exist, indicate the boundaries, $\mu=\pm\mu_c=\pm\sqrt{h^2-\Delta^2}$, of the topological phase in the thermodynamic limit.  Finally, the horizontal dashed lines highlight the finite-size energy levels at the critical points as predicted by theory through Eqs.~\eqref{v} and \eqref{Ediscrete}.}
\label{fig:muScans}
\end{figure*}

\subsection{Chemical potential scans at fixed Zeeman field} \label{sec:muScans}

Following measurements in Ref.~\onlinecite{Mingtang}, it is also interesting to contemplate finite-size effects at the topological phase transition exhibited by Eq.~\eqref{H} upon varying the chemical potential, $\mu$, and hence the electron density, $n$, at \emph{fixed} magnetic fields.  

We first briefly review the structure of the free-fermion band structure of Eq.~\eqref{H} at $\Delta=0$.  For concreteness we focus on $h < m\alpha^2$---the case of interest in the strong spin-orbit regime (see inset of Fig.~\ref{fig:StrongSO_FieldScans}, middle panel).  Here the wire has no electrons when $\mu < \mu_\mathrm{bottom} = m\alpha^2/2 - h^2/(2m\alpha^2)$; hosts two sets of Fermi points for $\mu_\mathrm{bottom} < \mu < -h$ or $\mu > h$; and most interestingly realizes a spinless regime with one pair of Fermi points for $-h < \mu < h$.  Let us now resurrect finite $\Delta$.  If $h>\Delta$ and the wire is sufficiently long, the system admits a topological phase within a subregion $-\mu_c < \mu < \mu_c$ of the spinless regime, where $\mu_c = \sqrt{h^2-\Delta^2}$.  We thus have topological phase transitions at critical values $\mu = \pm\mu_c$.  At these critical points, our analysis of the spectra for finite-size systems from Sec.~\ref{sec:FiniteSize} (see also Appendix \ref{FieldDependenceAppendix}) carries over completely.

To expose the resulting physics, Fig.~\ref{fig:muScans} presents the numerically calculated end-of-wire LDOS at fixed $h/\Delta$ versus $\mu/\Delta$. We again focus on the strong spin-orbit regime ($m\alpha^2/\Delta = 14$) for a relatively `short' wire ($L/\ell = 5$); cf.~Fig.~\ref{fig:StrongSO_FieldScans}, left panel.  At zero field ($h/\Delta=0$), the levels form shifted parabolas with minimum energy $\Delta$.  This structure reflects a combination of $(i)$ the momentum quantization in our finite-size wire and $(ii)$ the quadratic energy dispersion---arising from the superconducting gap---near the Fermi wavevectors.  The latter vary approximately linearly with $\mu$ over sufficiently short intervals centered at the bottom of the parabolas\footnote{Taking periodic boundary conditions for simplicity, the final expression for a parabola centered at $\mu=\mu_0$ reads $E = \Delta + \frac{\hbar^2\beta^2}{2m_\mathrm{eff}}\left(\mu - \mu_0\right)^2$.  Here, $\mu_0$ is the chemical potential corresponding to the finite-size discrete Fermi wavevector $k_{F0}=2\pi n_0/L$ closest to the infinite system Fermi wavevector $k_F$, $m_\mathrm{eff}$ is the effective mass of low-energy excitations above $\Delta$, and $\beta$ is the proportionality factor relating small deviations in $k_F$ about $k_{F0}$ to small deviations in $\mu$ about $\mu_0$:  $k_F\approx k_{F0} + \beta(\mu-\mu_0)$.}. 
For long wires, we recover a continuum of states above energy $\Delta$ with a level spacing which decreases as $1/L$.  In modeling an experimental system such as those in Refs.~\onlinecite{Marcus15_NatureNano_10_232, Mingtang}, it is thus reasonable to interpret our approach here as putting in by hand the energy $\Delta$ of the lowest-lying (extended in our model) Andreev bound states at zero field arising from a more exact treatment of the proximity effect\footnote{See, for example, Ref.~\onlinecite{Mutation}.  In that work, it was reported that even at zero field the pair of lowest-lying, spin-degenerate Andreev bound states may actually be localized to the ends of the wire due to finite spin-orbit coupling.}.  For eventual Majorana physics at finite $h$, it has been argued\cite{Cole15_PRB_92_174511_TooMuchGood} that this $\Delta$ should only be a fraction of the parent superconductor's gap giving rise to proximity-induced pairing.  With the above interpretation in mind, this indeed appears to be the case in the zero-field data of Ref.~\onlinecite{Mingtang}.  

As we turn on the Zeeman field, a state centered at $\mu=0$ begins to descend.  This 
state eventually evolves into a Majorana zero mode once in the topological phase.  For $h < \Delta$, there is no topological phase for any value of $\mu$, while for $h > \Delta$, the (infinite system) topological regime in the interval $-\mu_c < \mu < \mu_c$ lies between the pair of vertical dashed lines in the bottom panels of Fig.~\ref{fig:muScans}.   (At $h/\Delta=1$, the topological phase shrinks to a point at $\mu_c=\sqrt{h^2-\Delta^2}=0$; this situation appears in the upper-right panel.)  For cases in which a topological regime exists, we also show the finite-size energy levels at the critical points $\pm\mu_c$ as given by Eqs.~\eqref{v} and \eqref{Ediscrete} via horizontal dashed lines as in Figs.~\ref{fig:StrongSO_FieldScans} and \ref{fig:WeakSO_FieldScans}.  The agreement between theory and numerics is quite good, especially for larger $h/\Delta$ where the zero modes span larger ranges of $\mu$.  Collectively, these plots further demonstrate the robustness of Majorana zero-mode physics in `short' wires with strong spin-orbit coupling.

In the weak spin-orbit regime, robust Majorana zero modes in the topological phase still require a sharp phase transition at both critical points $\mu=\pm\mu_c$, and the level structure given by  Eq.~\eqref{Ediscrete} still applies.  However, for weak spin-orbit coupling, the phase transitions on the low- and high-density sides can differ in terms of visibility for end-of-wire conductance probes.  Notably, the low-lying states on the trivial low-density side have very poor visibility, while on the trivial high-density side a pair of end-of-wire localized Andreev bound states appear that are energetically separated from the bulk states.  Upon entering the topological phase, one of these Andreev bound states becomes the $n=0$ level that evolves into Majorana modes while the other becomes the $n=1$ bulk mode, with energies at the transition still given by Eq.~\eqref{Ediscrete}.

Finally, we have thus far only considered the situation with a constant pairing amplitude $\Delta$.  It is well known, however, that $\Delta$ renormalizes downwards as the field increases.  This effect is particularly important for Al/InAs experiments in, for example, Refs.~\onlinecite{Marcus15_NatureNano_10_232, Mingtang} where the critical magnetic field of the parent superconductor $B_{\rm SC}$ may only be a factor of two or three greater than the magnetic field $B_c$ at the topological phase transition itself.  In Appendix~\ref{PairingSuppressionAppendix}, we include the effects of reasonable pairing suppression due to the Zeeman field and show that the results in Figs.~\ref{fig:StrongSO_FieldScans} and \ref{fig:muScans} remain qualitatively intact, albeit with some quantitative differences.

\section{Experimental protocols} \label{sec:FSSprotocol}

With the advent of clean, hard-gap wires \cite{Marcus15_NatureMat_14_400, Marcus15_NatureNano_10_232} that largely eliminate unwanted background conductance, mapping the low-lying level structure in fields should be possible through NS tunneling spectroscopy \cite{mourik12,das12,finck12,Churchill}.  In fact, our study was strongly informed by recent measurements that clearly resolve multiple sub-gap states---including a level that evolves into a zero-bias peak\cite{Mingtang}.  One relatively simple way to quantify the approach of a topological phase transition in such experiments is through the field or density dependence of the finite-size energy levels \emph{near} the critical point [e.g., the behavior of the bottom two levels near the vertical dashed line in Fig.~\ref{fig:StrongSO_FieldScans}(a)].  Conveniently, this characterization relies on measurements of a system with fixed length $L$ that can be sufficiently small that the transition supports very few sub-gap states.  

We describe in Appendix \ref{FieldDependenceAppendix} a very general procedure for analytically tracking the evolution of the system's energy levels as a function of a tuning parameter---such as magnetic field or chemical potential---in the vicinity of the topological phase transition.  While the procedure should be applicable to any microscopic model, e.g., one with multiple bands, we here specialize it to Eq.~\eqref{H}.  In this section, we further focus on the scenario where the phase transition occurs at $\mu\approx0$, an experimentally relevant limit which gives particularly simple and universal expressions for the energy levels.  In practice one can tune to this $\mu\approx0$ limit by adjusting side-gate voltages to minimize the Zeeman field necessary to produce Majorana-induced zero-bias peaks.  Transitions at finite $\mu$ yield energies that are more complicated yet easily obtainable as outlined in Appendix \ref{FieldDependenceAppendix}.
 
 \begin{figure}[t]
\centering{
\subfigure{\includegraphics[width=0.7\columnwidth]{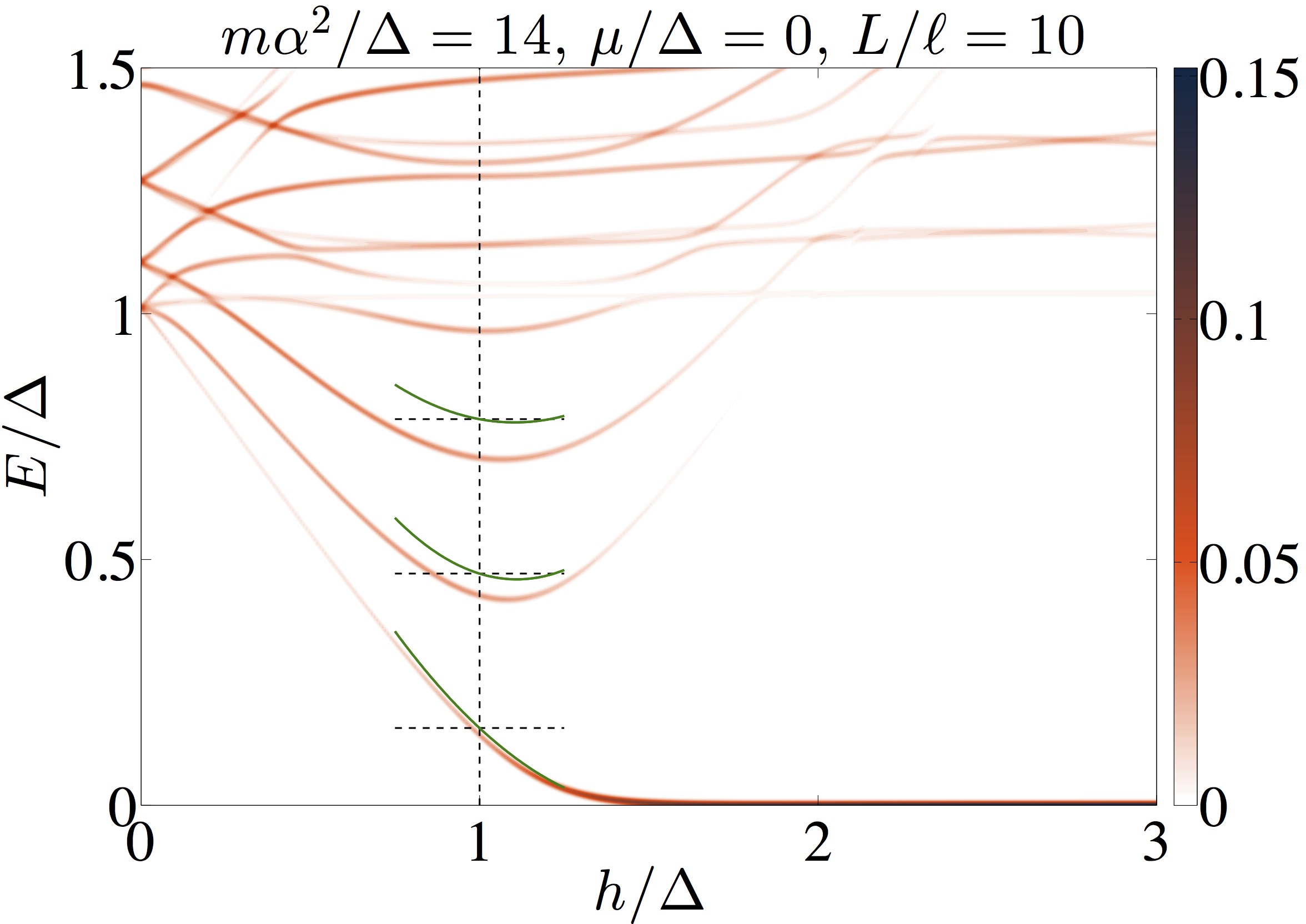}}
\subfigure{\includegraphics[width=0.7\columnwidth]{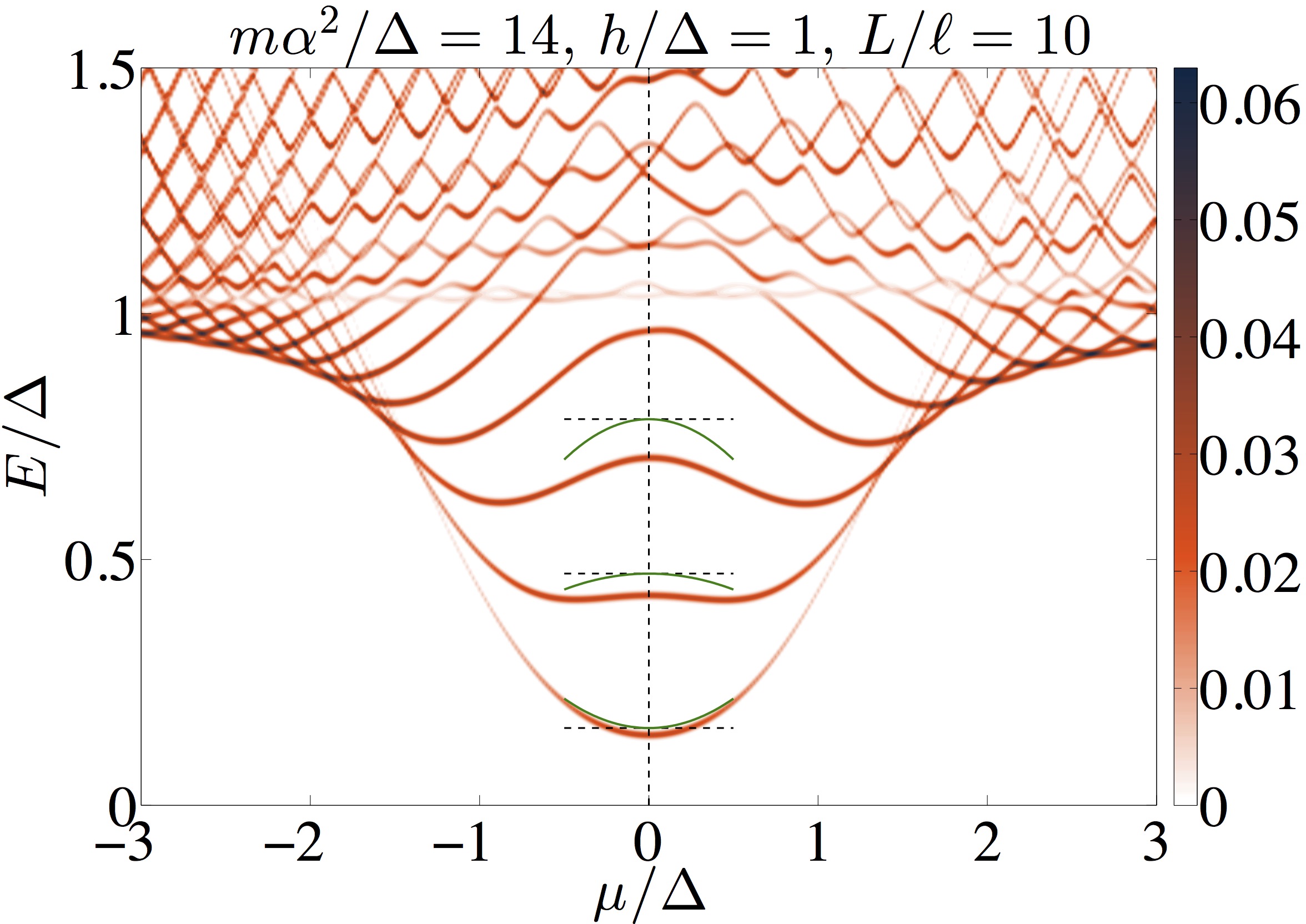}}
}
\caption{Level tracking at the critical point:  LDOS versus $h/\Delta$ and $E/\Delta$ at $\mu=0$ (top panel; cf.~Fig.~\ref{fig:StrongSO_FieldScans}) and $\mu/\Delta$ and $E/\Delta$ at $h=\Delta$ (bottom panel; cf.~Fig.~\ref{fig:muScans}) in the strong spin-orbit regime, $m\alpha^2/\Delta=14$, on a system of length $L/\ell=10$.  These plots demonstrate the accuracy of the analytical level-tracking formulas, Eqs.~\eqref{En_B} and \eqref{En_mu}, presented in the text (solid green curves) for the lowest-lying three levels ($n=0,1,2$) at the transition.}
\label{fig:LevelTracking}
\end{figure}
 
For a system at constant $\mu=0$ with variable magnetic field $B$ (cf.~Figs.~\ref{fig:StrongSO_FieldScans} and \ref{fig:WeakSO_FieldScans}), we find that for fields close to the critical magnetic field $B_c$, the discrete energies in Eq.~\eqref{Ediscrete} are modified to
\begin{equation}
  \mathcal{E}_n \approx \frac{\pi\hbar\alpha}{L}\left(n + \frac{1}{2}\right) - c_{n1} \frac{1}{2}g\mu_B\delta B + c_{n2}\frac{\left[\frac{1}{2}g\mu_B\delta B\right]^2}{\frac{\pi\hbar\alpha}{L}\left(n + \frac{1}{2}\right)},
  \label{En_B}
\end{equation}
an expression good to $O(\delta B^2)$ where $\delta B = B - B_c$.  The numerical factors $c_{n1,2}$ are given in Eq.~\eqref{cn}.  This form assumes that the magnetic field alters the Zeeman energy but not the pairing potential $\Delta$ or chemical potential.  
The levels depend on parameters $g$, $B_c$, and $\alpha/L$.  One can roughly estimate $g$ through the slope of the lowest-lying level $\mathcal{E}_{n = 0}$ on the trivial side of the transition; $B_c$ through the field that minimizes $\mathcal{E}_{n = 1}$; and $\alpha/L$ from the measured difference $\mathcal{E}_{n = 1} - \mathcal{E}_{n = 0}$ at $B_c$.  These rough estimates can be refined through a more careful fit to Eq.~\eqref{En_B} for the lowest experimentally resolved sub-gap levels (whose properties should be most accurately captured by this expression).  Importantly, if $L$ is known then such fits yield the spin-orbit strength $\alpha$.  

In the top panel of Fig.~\ref{fig:LevelTracking}, we overlay the predictions of Eq.~\eqref{En_B} on top of the numerical data from Fig.~\ref{fig:StrongSO_FieldScans} at the intermediate length $L/\ell=10$ for the lowest three levels ($n=0,1,2$).  We see that the agreement is quite good given the discrepancy already apparent at $\delta B=0$ on this system size (see Fig.~\ref{ScalingFig}).  The quantitative agreement between Eq.~\eqref{En_B} and the numerics continues to improve as we increase $L$ (not shown), thereby confirming the validity of the approach spelled out in Appendix~\ref{FieldDependenceAppendix}.

It would also be interesting to tune through the topological phase transition by changing $\mu$ via nearby side-gate voltages\footnote{See Ref.~\onlinecite{Akhmerov_ElectrostaticMajorana} for a recent in-depth discussion on the effect of the electrostatic environment on Majorana nanowire hybrid devices}, although obtaining such large sweeps as in Fig.~\ref{fig:muScans} may be practically difficult.  Tracking of the finite-size levels near the critical values $\mu=\pm\mu_c=\pm\sqrt{h^2-\Delta^2}$ on a single, fixed-length device is possible here too.  We again focus on the experimentally relevant case of $\mu_c\approx0$ which occurs for fields $h\approx\Delta$.  Taking $\mu_c=0$, $h=\Delta$ exactly (cf.~Fig.~\ref{fig:muScans}, upper-right panel) so that the topological phase has shrunk to a point upon scanning $\mu$, we find that the energy levels near the transition evolve as
\begin{equation}
  \mathcal{E}_n \approx \frac{\pi\hbar\alpha}{L}\left[1-\frac{\delta\mu^2}{2\Delta^2}\right]\left(n + \frac{1}{2}\right) + c_{n1}\frac{\delta\mu^2}{2\Delta},
  \label{En_mu}
\end{equation}
which is good to $O(\delta\mu^2)$ where $\delta\mu = \mu - \mu_c = \mu$.  In the bottom panel of Fig.~\ref{fig:LevelTracking}, we overlay the levels from Eq.~\eqref{En_mu} onto the corresponding numerical data, again at $L/\ell=10$.  The agreement is reasonable and indeed improves as we increase $L$.  Note that the curvature of the levels decreases with increasing $n$.  Similarly to the field-scan protocol discussed above, it would be worthwhile to fit predictions such as Eq.~\eqref{En_mu} to the experimental data.

Useful device information can be extracted from a fixed-length wire even away from the $\mu \approx 0$ limit considered above.  For instance, one can quite generally estimate $B_c$ (and $h_c$ if the $g$-factor is known) by reading off the field that minimizes the $n=1$ level.  We point out that for short, strong-spin-orbit systems this method provides a more reliable way of determining $B_c$ than by the field at which zero modes form\cite{Loss13_PRB_87_024515}$^,$\footnote{However, field renormalization of the pairing gap [$\Delta=\Delta(h)$] as discussed in Appendix~\ref{PairingSuppressionAppendix} may render the $n=1$ state difficult to resolve\cite{Mingtang}.} (see Fig.~\ref{fig:StrongSO_FieldScans}, left panel).  
Assuming $\Delta$ is inferable from tunneling spectroscopy\footnote{This could be achieved, for example, through zero-field conductance measurements.  If desired renormalization of $\Delta$ by the field can be readily incorporated if the parent superconductor's critical field is known.}, identifying $h_c = \sqrt{\Delta^2+\mu^2}$ correspondingly determines $\mu$---which may be otherwise difficult to estimate.  Reading off $E_\mathrm{finite-size~gap}$ and using Eq.~\eqref{Egap1} with $v = \alpha \Delta/h_c$, one can then estimate $\alpha$---or more conservatively $\alpha/L$---at different gate voltages.  Even more simply, identifying $E_\mathrm{finite-size~gap}$ at a single, fixed gate voltage gives a very useful \emph{lower bound} on the spin-orbit strength since $\alpha\geq v=\frac{2LE_\mathrm{finite-size~gap}}{3\pi\hbar}$.  Interestingly, the only parameter required to establish this lower bound is the length $L$.  

As a more ambitious experiment, one could experimentally implement finite-size scaling to probe the predicted evolution of states sketched in Fig.~\ref{fig:StrongSO_FieldScans} as $L$ varies.  The bottom panels illustrate one possible way to perform the experiment using a single device.  Here a wire (e.g., InAs) is coated with superconducting islands (e.g., Al) separated by gate-tunable `valves'\cite{Gatemon,Gatemon2,Milestones} that control the coupling between adjacent islands.  The conductance is measured by sending in current from the normal lead to the leftmost island, which is grounded.  Successively opening and closing valves as in the figure effectively changes the length $L$ of the region probed by the lead and allows one to track the finite-size energy levels at the transition.  Observing the characteristic $1/L$ scaling of the bulk gap---i.e., $E_{n = 1}$---would provide additional sharp evidence for the expected critical behavior.  Here too such measurements constrain the system's Rashba spin-orbit coupling by virtue of Eqs.~\eqref{v} and \eqref{Egap1}.  As an independent check, $\ell = \hbar \alpha/\Delta$ can be separately inferred from the (exponential) $L$ dependence of the Majorana mode splitting that occurs in the topological phase.

Another strategy using more traditional technology would be to take a single, long superconducting island as in the device in Ref.~\onlinecite{Mingtang}, but place \emph{several} side gates of known lengths nearby.  Selectively tuning the side-gate voltages into and out of the topological regime systematically alters the length of the topological segment of the wire, thereby effectively changing $L$.  (A similar gate setup was already realized in the original Delft experiment\cite{mourik12}.)  We believe both schemes are quite reasonable with present technology and hope that they may be pursued in the near future.

Finally, we remark that the issue of poor visibility of bulk states in end-of-wire tunneling measurements \cite{StanescuTransition,Pientka,Prada} is expected to be alleviated in small systems for which the phase transition is very much a crossover.  Variations wherein tunneling occurs in the middle of the wire would, however, skirt this issue entirely.

\section{Discussion}

With the field of Majorana nanowires presently poised to move beyond zero-mode detection, we have examined several fundamental questions regarding the putative topological phase transition that accompanies the formation of these zero modes.  We have shown that---rather surprisingly---robust zero modes can easily exist in systems too small to exhibit anything close to a true topological phase transition, provided that the spin-orbit coupling is strong.  However, it should be emphasized that here there is no parametric suppression of the Majorana splitting once the finite-size gap at the phase transition becomes comparable to the induced pairing gap.  Numerical factors instead conspire to make this splitting quite small in practice.  In fact, there may actually be as few as two sub-gap states visible in the conductance spectra even if robust Majorana-induced zero-bias peaks appear in the topological phase.  This point is particularly noteworthy in light of recent experimental data on epitaxial Al/InAs hard-gap devices from Ref.~\onlinecite{Mingtang} which indeed may reside in this regime.

Since the original Majorana nanowire proposals in Refs.~\onlinecite{1DwiresLutchyn,1DwiresOreg}, there have been a proliferation of papers over the past several years on Eq.~\eqref{H} and its refinements, yet the fate of the first-excited state $n=1$ level on finite-size systems has gone relatively unexplored.  Since this is the lowest-lying level which is gapped on \emph{both} sides of the topological phase transition, 
understanding its behavior is of fundamental importance, especially when discussing physics of the phase transition itself.  We hope that our work will help to bring investigation of such sub-gap states into greater prominence.  Indeed studying the behavior of these levels as we have proposed in this work---on both fixed- and variable-length finite-size wires---can give very valuable universal information about the eventual topological phase transition expected at $L\to\infty$.  Furthermore, such analysis can provide nontrivial information about parameters of the hybrid device, most notably its effective spin-orbit coupling strength $\alpha$.  Experimentally probing the finite-size scaling and tracking of finite-size energy levels at the topological phase transition in Majorana nanowires thus constitutes a worthwhile pre-braiding endeavor in the Majorana problem.

\section*{Acknowledgments}
We are indebted to M.~Deng, K.~Flensberg, L.~Glazman, M.~Hell, and C.~Marcus for helpful discussions on this work, and especially thank M.~Deng and C.~Marcus for sharing unpublished results.  We also gratefully acknowledge support from the NSERC PGSD program (D.~A.); the National Science Foundation through grant DMR-1341822 (J.~A.); the Alfred P.\ Sloan Foundation (J.~A.); the Caltech Institute for Quantum Information and Matter, an NSF Physics Frontiers Center with support of the Gordon and Betty Moore Foundation through Grant GBMF1250; and the Walter Burke Institute for Theoretical Physics at Caltech.  Part of this work was performed at the Aspen Center for Physics, which is supported by National Science Foundation grant PHY-1066293 (R.~V.~M.).

\appendix

\section{Derivation of finite-system energies at the topological phase transition}
\label{TransitionAppendix}

\begin{figure}[t]
\includegraphics[width=\columnwidth]{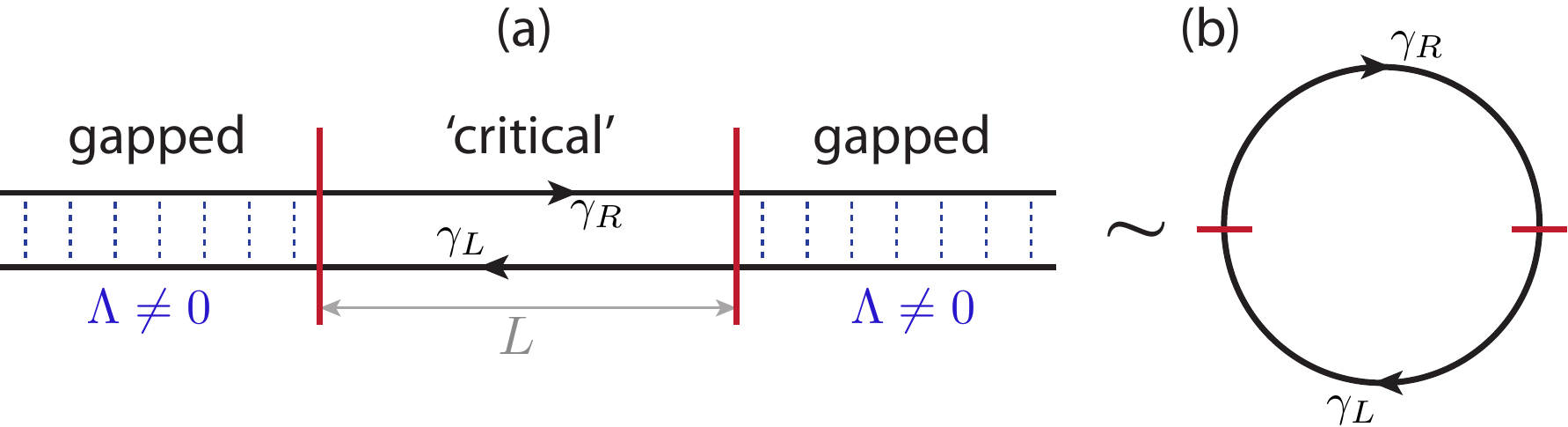}
\caption{(a) Schematic of the effective Hamiltonian in Eq.~\eqref{HeffL} describing a length-$L$ wire tuned to the critical point between topological and trivial phases.  Right- and left-moving Majorana fields $\gamma_{R/L}$ remain uncoupled in the central region but are gapped out by a hybridization $\Lambda$ elsewhere.  (b) In the $\Lambda \rightarrow \infty$ limit the system maps to the chiral edge of a $p+ip$ superconductor.}
\label{Critical_fig}
\end{figure}

\begin{figure}[b]
\includegraphics[width=3in]{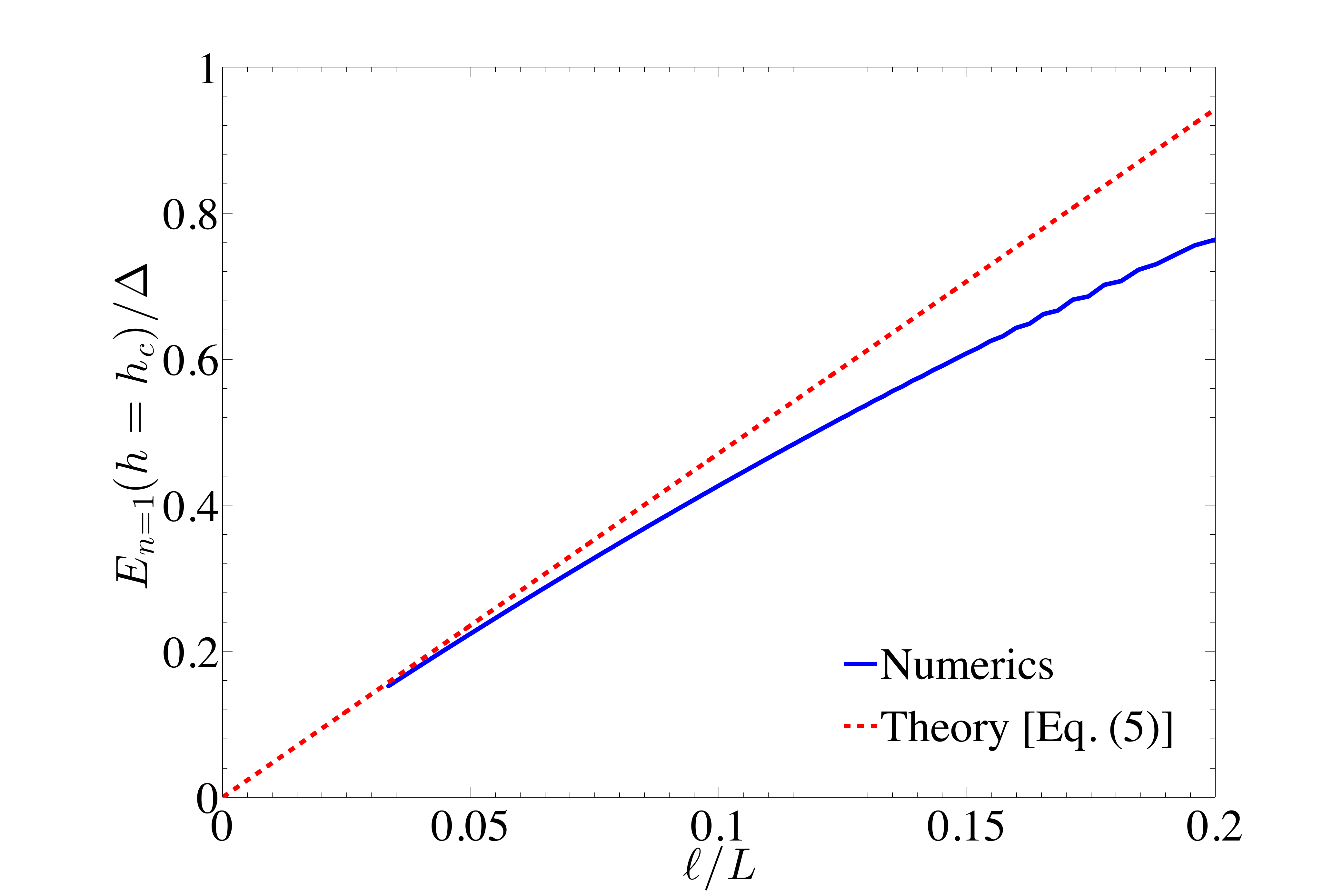}
\caption{Bulk gap $E_{n = 1}$ in units of $\Delta$ at the critical Zeeman strength $h = h_c$ with chemical potential $\mu = 0$.  Solid and dashed curves respectively correspond to numerical results and analytical predictions, while the horizontal axis represents $\ell/L = \hbar \alpha/(\Delta L)$.  As the length increases, the two curves nicely converge.}
\label{ScalingFig}
\end{figure}

This Appendix derives an effective low-energy Hamiltonian from Eq.~\eqref{H} at the critical magnetic field $h = h_c$, in particular to assess the influence of finite-size effects on the spectrum.  In an infinite system the energies at criticality are given by $E_k = \hbar v |k|$ [Eq.~\eqref{Ek}] with momentum $k$ a continuous parameter.  Projection onto these low-lying excitations follows by sending
\begin{eqnarray}
  \psi_\uparrow &\rightarrow& \frac{1}{\sqrt{2}}(-i e^{i \theta}\gamma_R + i e^{-i \theta}\gamma_L)
  \nonumber \\
  \psi_\downarrow &\rightarrow& \frac{1}{\sqrt{2}}(e^{-i \theta}\gamma_R + e^{i \theta}\gamma_L),
  \label{expansion}
\end{eqnarray}
where $\gamma_{R/L}$ are right/left-moving gapless Majorana fields and $\tan(2\theta) = \mu/\Delta$.  
The following elegant effective Hamiltonian for the transition then arises:
\begin{equation}
  H_{\rm eff} = \int_x \left(-i \hbar v \gamma_R\partial_x\gamma_R + i \hbar v \gamma_L \partial_x \gamma_L \right).
\end{equation}

\begin{figure*}
\centering{
\subfigure{\includegraphics[width=0.315\textwidth]{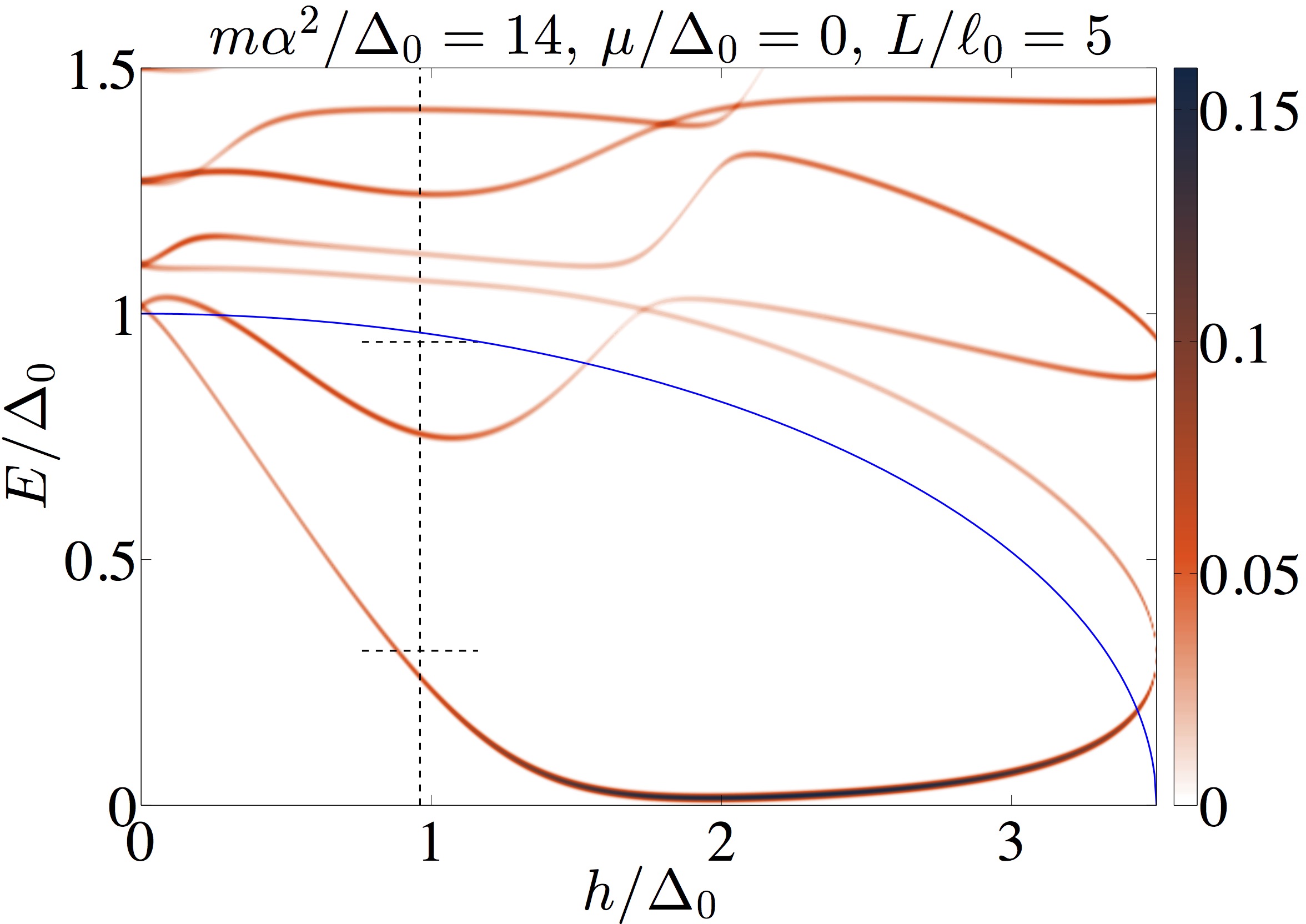}} \hfill
\subfigure{\includegraphics[width=0.315\textwidth]{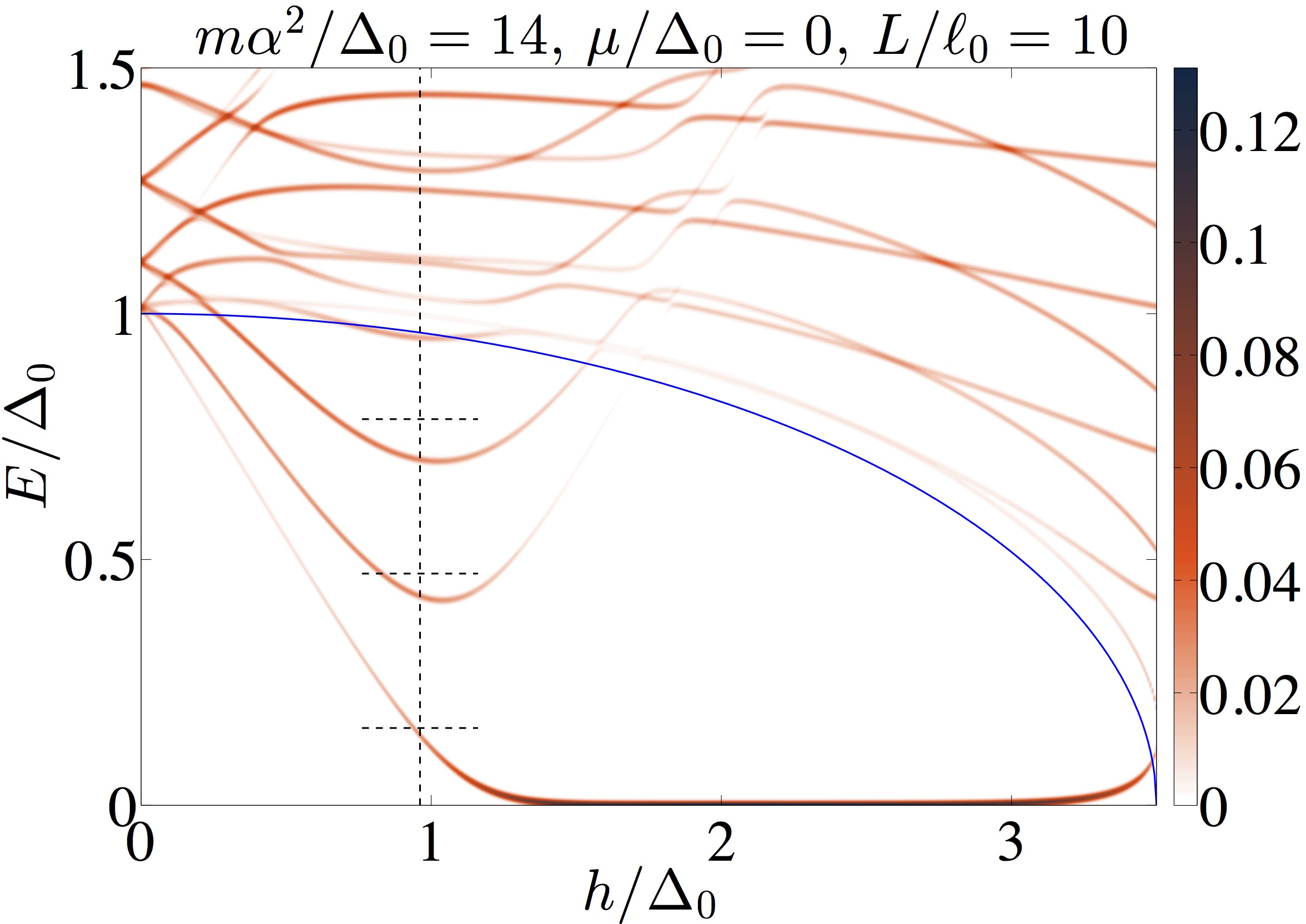}} \hfill
\subfigure{\includegraphics[width=0.315\textwidth]{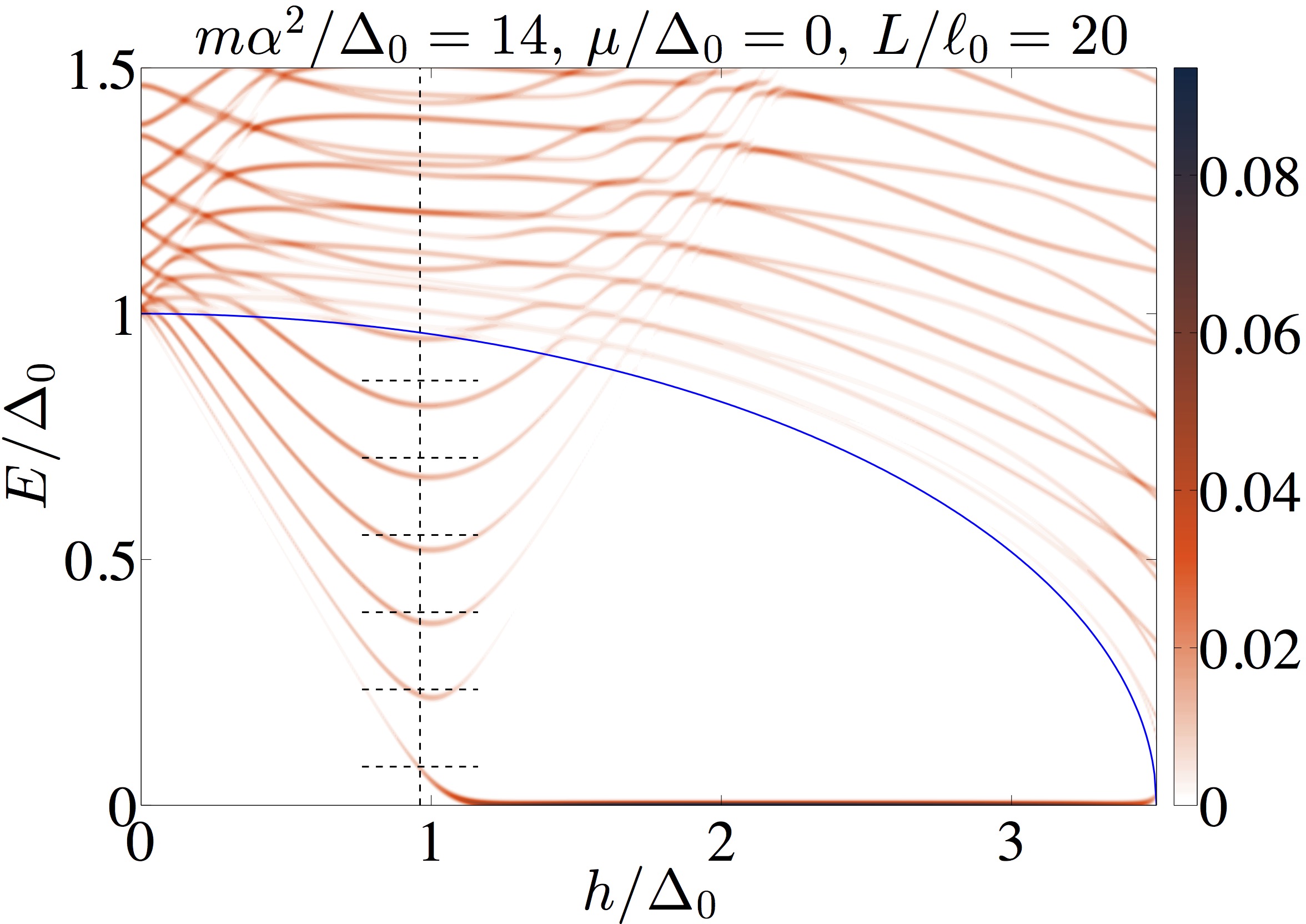}}
}
\caption{Zeeman field scans at strong spin-orbit coupling with a field-suppressed induced gap:  LDOS versus $h/\Delta_0$ and $E/\Delta_0$ with $\mu=0$ in the strong spin-orbit coupling regime, $m\alpha^2/\Delta_0=14$, including a field-suppressed superconducting gap given by Eq.~\eqref{DeltaOfh} with $h_{\rm SC} = 3.5\Delta_0$; we show $\Delta(h)$ as a solid blue curve in all plots.  As in Fig.~\ref{fig:StrongSO_FieldScans}, different panels correspond to system sizes $L/\ell_0=5,10,20$ from left to right, where here $\ell_0 \equiv \hbar\alpha/\Delta_0$.  The dashed vertical and horizontal lines carry the same meaning as in Figs.~\ref{fig:StrongSO_FieldScans} and \ref{fig:WeakSO_FieldScans}, now taking into account Eq.~\eqref{DeltaOfh}.
}
\label{fig:StrongSO_FieldScans_DeltaSup}
\end{figure*}

The spectrum for a finite-size system of length $L$ can be efficiently derived by modifying the low-energy Hamiltonian above to
\begin{equation}
  H_{\rm eff} \rightarrow \int_x \left[-i \hbar v \gamma_R\partial_x\gamma_R + i \hbar v \gamma_L \partial_x \gamma_L  + 2i \Gamma(x) \gamma_R \gamma_L\right]
  \label{HeffL}
\end{equation}
where
\begin{equation}
  \Gamma(x)=\left\{
                \begin{array}{ll}
                  0, & |x| < L/2 \\
                  \Lambda > 0, & |x|>L/2
                \end{array}
              \right.
\end{equation}
introduces a boundary to the `critical' wire by gapping the adjacent regions; see Fig.~\ref{Critical_fig}(a).  Solving for the wavefunctions in each piecewise-uniform region and matching boundary conditions yields quantized momenta $k_n = \frac{\pi}{L}(n+\frac{1}{2})$ in the $\Lambda \rightarrow \infty$ limit ($n$ is an integer).  One can intuitively understand this result as follows.  Figure~\ref{Critical_fig}(b) illustrates that the system maps to a single chiral Majorana fermion on a ring of circumference $\tilde L = 2L$---precisely as in the edge of a two-dimensional spinless $p+ip$ superconductor.  The chiral fermion must exhibit anti-periodic boundary conditions, since periodic boundary conditions would yield a single Majorana zero mode with no partner (which is impossible).  From this perspective the momenta are immediately given by $k_n = \frac{2\pi}{\tilde L}(n+\frac{1}{2})$, in harmony with the result quoted above.  

Inserting these quantized momenta into the continuum energy $E_k = \hbar v |k|$ yields the discrete spectrum specified in Eq.~\eqref{Ediscrete}---in particular with a finite-size bulk gap $E_{n = 1} =  \frac{3\pi \hbar v}{2L}$ [Eq.~\eqref{Egap1}] corresponding to the $k_{n = 1}$ mode.  (As noted in the main text the $k_{n = 0}$ mode is special because it evolves into a Majorana zero mode on the topological side of the transition.)  It is useful to systematically compare the $E_{n = 1}$ bulk gap derived from the low-energy effective Hamiltonian with that obtained numerically from the more microscopic model of Eq.~\eqref{H}.  Figure~\ref{ScalingFig} illustrates the length dependence for these analytical and numerical values using the same parameters as for Fig.~\ref{fig:StrongSO_FieldScans}.  Both figures show that the analytical result indeed converges well with numerics as the system size increases.

\section{Field and chemical potential dependence of finite-size energies near the topological phase transition}
\label{FieldDependenceAppendix}

Our goal here is to extend the analysis from Appendix \ref{TransitionAppendix} to extract the finite-size energy levels for systems tuned slightly away from criticality.  Sufficiently close to the transition, the structure of the energies is expected to remain universal and well-captured by the effective Hamiltonian in Eq.~\eqref{HeffL}.  To model the system off criticality we now take
\begin{equation}
  \Gamma(x)=\left\{
                \begin{array}{ll}
                  M, & |x| < L/2 \\
                  \Lambda > 0, & |x|>L/2
                \end{array}
              \right. ,
\end{equation}
where we have added a mass $M$ coupling right- and left-moving Majorana fields in the wire region of length $L$.  We will again take $\Lambda \rightarrow \infty$ in the adjacent outer regions to impose hard-wall boundary conditions on the wire.  A positive mass $M>0$ (corresponding to the same sign mass in the wire and outer regions) moves the system off criticality into the trivial state; for $M<0$ (corresponding to opposite-sign masses) the topological phase instead appears.  

Eigenstates of the effective Hamiltonian can again be obtained straightforwardly by solving for the wavefunctions in each piecewise-uniform region and imposing boundary conditions.  Carrying out this exercise, we find that the energies $\mathcal{E}$ must satisfy 
\begin{equation}
  e^{i 2 k L} = \left(\frac{1+i A}{1-iA}\right)^2
  \label{EnergyEq}
\end{equation}
with
\begin{equation}
  k = \frac{1}{v}\sqrt{\mathcal{E}^2-M^2},~~~~~A = \frac{\sqrt{\mathcal{E}^2-M^2}}{\mathcal{E}-M}.
\end{equation}
When $M = 0$ the energies are
\begin{equation}
  E_n = \hbar v k_n,~~~~k_n = \frac{\pi}{L}\left(n+\frac{1}{2}\right)
\end{equation}
for non-negative integers $n$, as obtained in Appendix~\ref{TransitionAppendix}.  Corrections arising from a finite mass $M$ may be obtained by assuming a power-series:
\begin{equation}
  \mathcal{E}_n = E_n + c_{n1} M + c_{n2} \frac{M^2}{E_n} + \cdots.
  \label{EEn}
\end{equation}
Inserting this ansatz into Eq.~\eqref{EnergyEq} yields
\begin{equation}
  c_{n1} = \frac{1}{\pi(n+1/2)},~~~~~c_{n2} = \frac{1}{2}-\frac{1}{[\pi(n+1/2)]^2}.
  \label{cn}
\end{equation}

\begin{figure*}
\centering{
\subfigure{\includegraphics[width=0.315\textwidth]{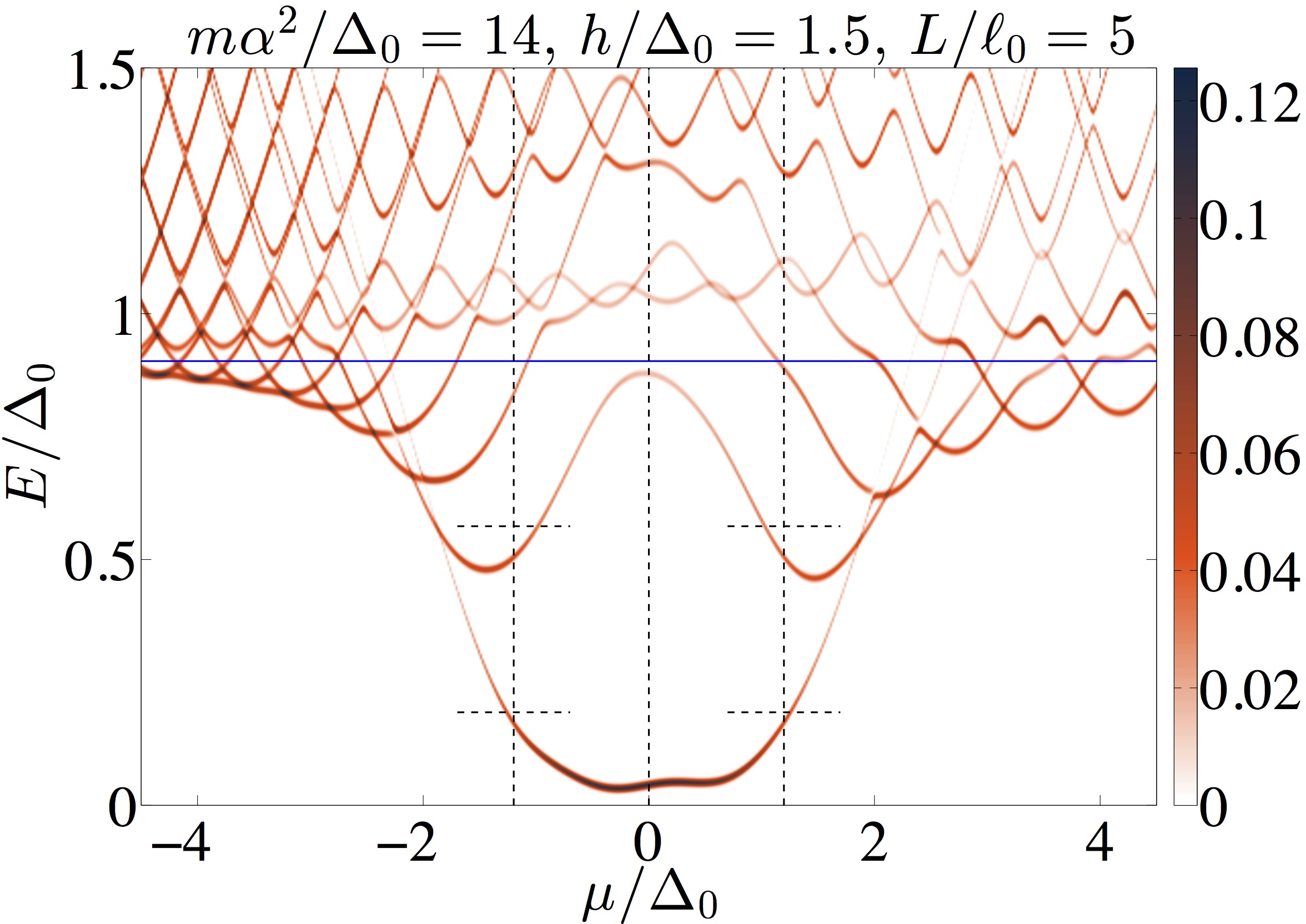}} \hfill
\subfigure{\includegraphics[width=0.315\textwidth]{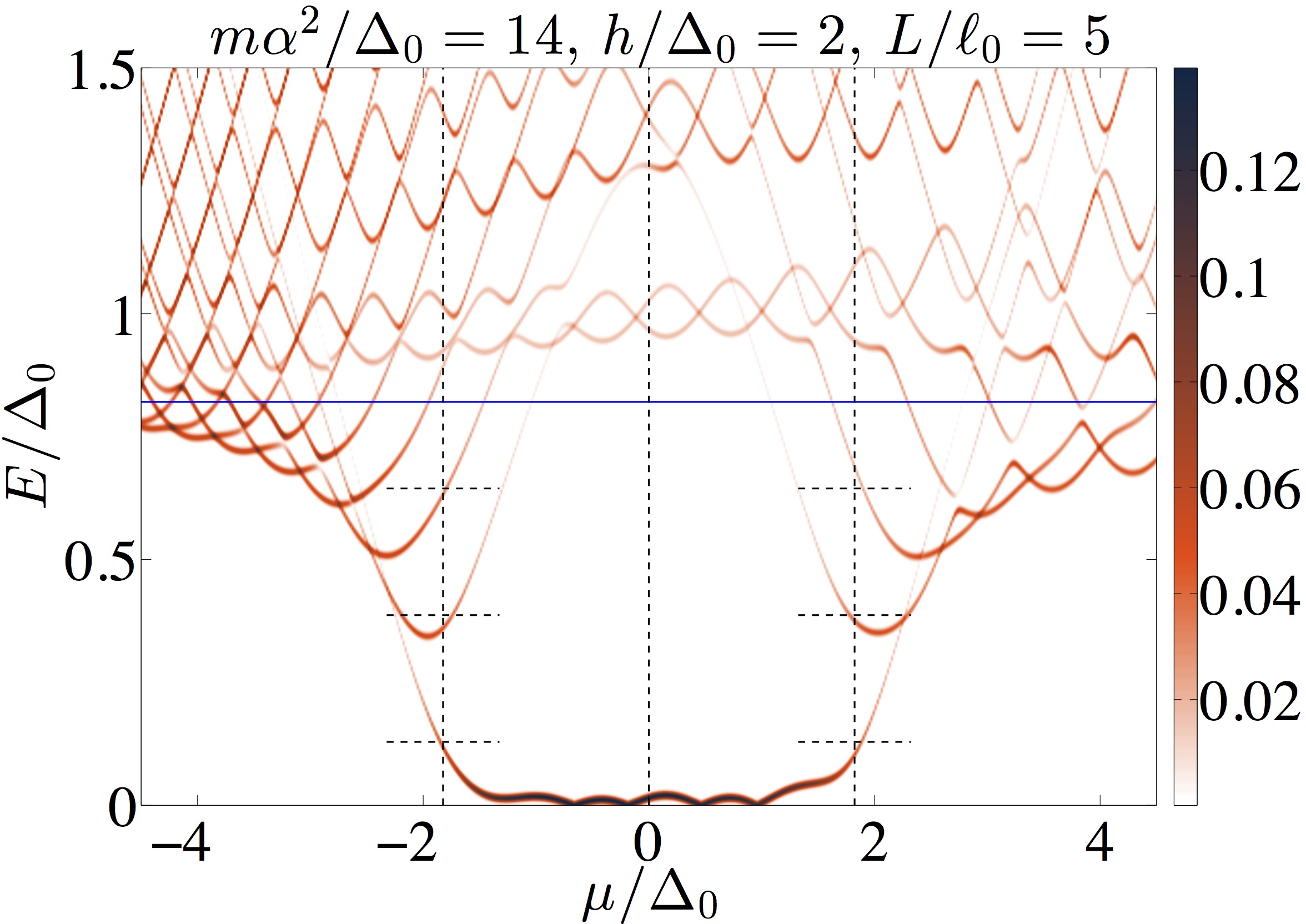}} \hfill
\subfigure{\includegraphics[width=0.315\textwidth]{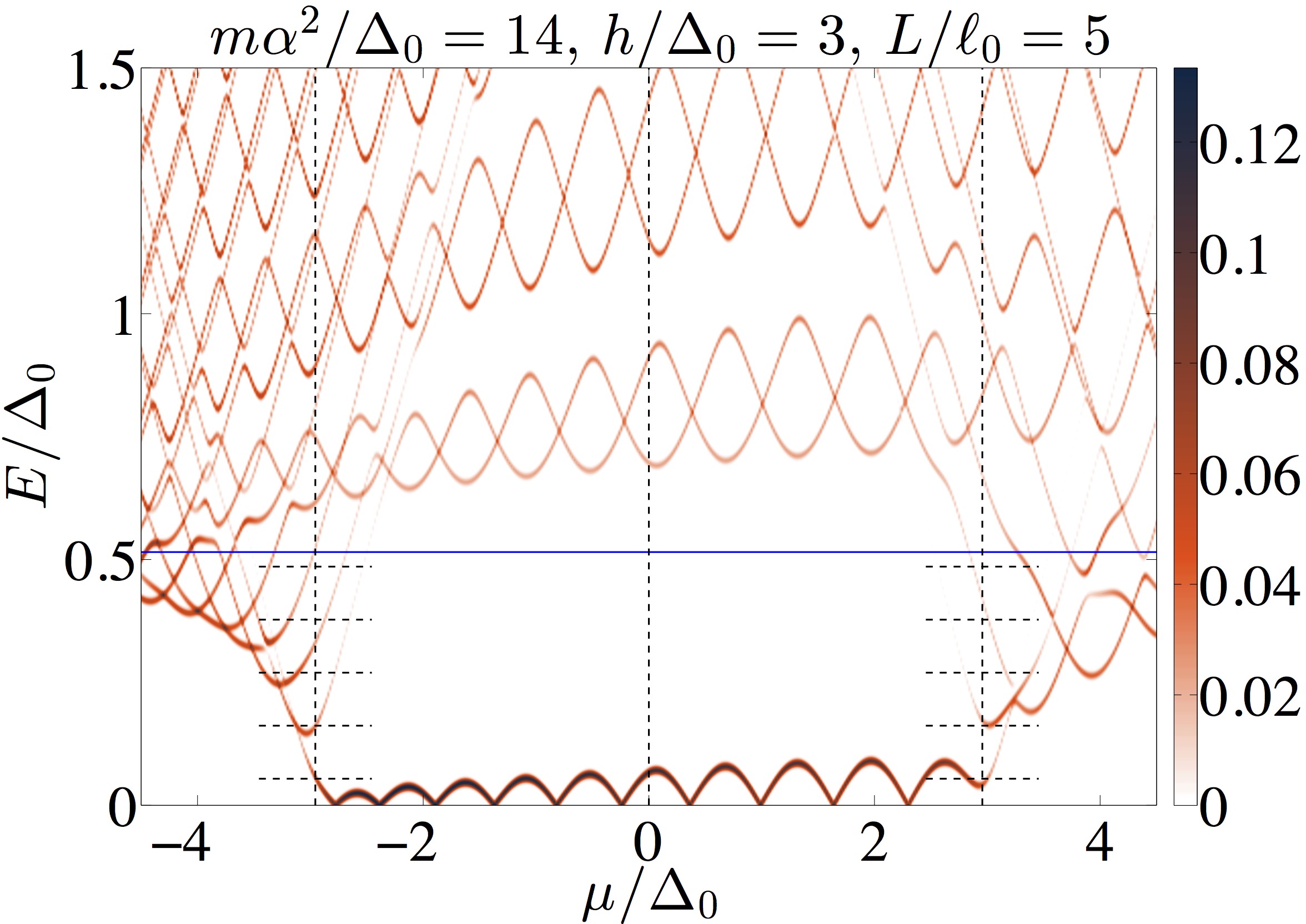}}
}
\caption{Chemical potential scans at strong spin-orbit coupling with a field-suppressed induced gap:  LDOS versus $\mu/\Delta_0$ and $E/\Delta_0$ on a system of length $L/\ell_0=5$ in the strong spin-orbit coupling regime, $m\alpha^2/\Delta_0=14$, including a field-suppressed superconducting gap given by Eq.~\eqref{DeltaOfh} with $h_{\rm SC} = 3.5\Delta_0$; the value of $\Delta(h)$ used in each plot is marked by a horizontal blue line.  As in Fig.~\ref{fig:muScans}, different panels correspond to different values of the Zeeman energy $h/\Delta_0$, and the meaning of the dashed vertical and horizontal lines is the same.}
\label{fig:StrongSO_muScans_DeltaSup}
\end{figure*}

These energies depend on the velocity $v$ (through $E_n$) and the mass $M$, both inputs to the effective model [Eq.~\eqref{HeffL}].
To relate these quantities to parameters in a given microscopic model such as Eq.~\eqref{H}, we can employ the following procedure.  We know that for a system with periodic rather than hard-wall boundary conditions, the energy dispersion of our effective Hamiltonian at finite $M$ reads
\begin{equation}
E_\mathrm{eff}(k) = \sqrt{(\hbar v k)^2 + M^2}.
\end{equation}
We can thus expand the square of the dispersion for our microscopic model, $E_\mathrm{micro}^2(k)$, about $k=0$ and identify the $O(k^2)$ term with $(\hbar v k)^2$ and the $O(k^0)$ term with $M^2$.  Applying this algorithm to Eq.~\eqref{H}, we find
\begin{equation}
v^2 = \frac{\mu}{m}\left[\frac{h}{\sqrt{\Delta^2+h^2}} - 1\right] + \alpha^2\left[1 - \frac{\mu^2}{h\sqrt{\Delta^2+\mu^2}} \right],
\end{equation}
\begin{equation}
M^2 = h^2 + \Delta^2 + \mu^2 - 2h\sqrt{\Delta^2+\mu^2}.
  \label{Meq}
\end{equation}

Next, we expand these expressions in a power series in the deviations away from the critical point (e.g., in $\delta h$ or $\delta\mu$) and insert the result into Eq.~\eqref{EEn} to derive leading-order level tracking formulas such as Eqs.~\eqref{En_B} and \eqref{En_mu} in the main text.  Expanding about a critical point with $\mu=0$ yields particularly simple results; see Eqs.~\eqref{En_B} and \eqref{En_mu} and Fig.~\ref{fig:LevelTracking}.  The procedure is, however, still valid at finite $\mu$; for instance, the energies so obtained describe well the low-lying levels near the critical points in the bottom panels of Fig.~\ref{fig:muScans}.

Finally, we note that Eq.~\eqref{Meq} determines $M$ only up to a sign.  The sign can be easily fixed, however: $M$ is positive (negative) if the tuning parameter takes the system into the trivial (topological) phase.

\section{Effect of pairing suppression by the magnetic field}
\label{PairingSuppressionAppendix}

Suppression of the pairing energy $\Delta$ by the magnetic field was so far ignored but can quantitatively effect the level structure over the field intervals displayed in Figs.~\ref{fig:StrongSO_FieldScans}, \ref{fig:WeakSO_FieldScans}, and \ref{fig:muScans}.  Such effects are, for example, present in the Al/InAs devices studied in Ref.~\onlinecite{Mingtang}.  We now incorporate pairing suppression by assuming a Zeeman field-dependent pairing amplitude
\begin{equation}
  \Delta(h) = \Delta_0\sqrt{1-\left(\frac{h}{h_{\rm SC}}\right)^2}
  \label{DeltaOfh}
\end{equation}
in our simulations of Eq.~\eqref{H}.  Here $\Delta_0$ is the induced pairing amplitude at zero field and $h_{\rm SC} = \frac{1}{2}g\mu_B B_{\rm SC}$ is the Zeeman energy associated with the parent superconductor's critical magnetic field $B_{\rm SC}$; $h_{\rm SC}$ should not be confused with the critical value $h_c$ at which the topological phase transition arises.  [In the main text, we took a field-independent $\Delta(h) = \Delta = \Delta_0$, corresponding to $h_\mathrm{SC}\to\infty$ in Eq.~\eqref{DeltaOfh}.]

We present in Fig.~\ref{fig:StrongSO_FieldScans_DeltaSup} scans of the end-of-wire LDOS versus Zeeman field as in Fig.~\ref{fig:StrongSO_FieldScans}, still at strong spin-orbit coupling with $m\alpha^2/\Delta_0=14$, but now with a field-renormalized induced pairing gap given by Eq.~\eqref{DeltaOfh} with $h_{\rm SC}=3.5\Delta_0$.  For units we use the zero-field gap $\Delta_0$ and the length scale $\ell_0\equiv\hbar\alpha/\Delta_0$.  Figure \ref{fig:StrongSO_muScans_DeltaSup} shows corresponding scans versus chemical potential at a few values of $h/\Delta_0$ for the shortest wire, $L/\ell_0=5$.

Overall, the physics is qualitatively similar to the constant-$\Delta$ results of Figs.~\ref{fig:StrongSO_FieldScans} and \ref{fig:muScans}.  The finite-size level structure at the eventual topological phase transition is basically unaffected.  However, now the zero mode in the shortest wire ($L/\ell_0=5$) does begin to experience noticeable splitting as we approach $h=h_{\rm SC}$.  Still, in this strong spin-orbit regime, this zero mode remains reasonably robust over an appreciable field and chemical potential range (see Fig.~\ref{fig:StrongSO_FieldScans_DeltaSup}, left panel, and Fig.~\ref{fig:StrongSO_muScans_DeltaSup}).  On the other hand, for the longer wires ($L/\ell_0=10, 20$) the zero modes remain intact essentially right up to $h = h_{\rm SC}$.

Finally, we point out the dense set of levels developing above $\Delta(h)$ in the topological regime for the larger sizes in Fig.~\ref{fig:StrongSO_FieldScans_DeltaSup}---strong spin-orbit coupling optimizes the excitation gap to very near $\Delta(h)$ as expected.  In a real experiment as in Ref.~\onlinecite{Mingtang}, however, we should additionally expect a continuum of states above the parent superconductor's gap $\Delta_{\rm parent}(h)$ for all $h$ even on the shortest wires.  Our simplified model in Eq.~\eqref{H} puts in proximity-induced pairing by hand and thus is unable to capture this experimental feature (see also the discussion in Sec.~\ref{sec:muScans}).  A more accurate modeling of the proximity effect should produce such a continuum of levels above $\Delta_{\rm parent}(h)$ in the left panel of Fig.~\ref{fig:StrongSO_FieldScans_DeltaSup}, as well as cause level repulsion between those parent-superconductor states and the all-important $n=1$ level for fields $h\gtrsim h_c$.  This all seems consistent with the experimental data from Ref.~\onlinecite{Mingtang}.

\section{Details of the numerical calculations}
\label{NumericalDetails}

We discretize the Bogoliubov-de Gennes (BdG) Hamiltonian, Eq.~\eqref{H}, into a tight-binding model of $N_\mathrm{sites}$ with lattice spacing $a = L/N_\mathrm{sites}$.  The corresponding discrete, sparse BdG Hamiltonian is then diagonalized exactly targeting the $\sim$50 states nearest zero energy using a shift-and-invert routine.

We define the local density of states (LDOS) as
\begin{equation}
\rho(E, x) = \sum_{i;\,s=\uparrow,\downarrow}\left[|u_i^s(x)|^2 + |v_i^s(x)|^2\right]\delta(E - E_i),
\label{eq:LDOS}
\end{equation}
where $u_i^s$ and $v_i^s$ are the eigenvectors of Eq.~\eqref{H} in the particle and hole sectors, respectively; the summation over $i$ includes all $4N_\mathrm{sites}$ eigenpairs of the discretized BdG Hamilitonian.  The $\delta$-function is regularized with a normalized Gaussian of width $\sigma$, i.e., $\delta(E) = \frac{1}{\sqrt{2\pi}\sigma} e^{-E^2/2\sigma^2}$, which in our calculations we have taken to be $\sigma = 0.005\Delta_0$.  For the end-of-wire LDOS as displayed in all figures, we plot $\rho(E, x)$ averaged over the leftmost 5\% of the wire.  The actual numerical value of the LDOS as shown in the plots is dependent on our normalization conventions, discretization, and $\delta$-function regularization, and it is thus not particularly important; yet it can be somewhat meaningfully compared across plots.

In all of our numerics, we have been careful to converge to the continuum limit by choosing $N_\mathrm{sites}$ sufficiently large.  Specifically, to avoid unwanted lattice effects we require $k_{F}a \ll \pi$, where $k_F$ is the (largest) Fermi wavevector in the free-fermion band structure.  For an infinite system we have
\begin{equation}
k_F = \frac{\sqrt{2m}}{\hbar}\sqrt{\mu + m\alpha^2 + \sqrt{(\mu + m\alpha^2)^2 + (h^2 - \mu^2)}}.
\label{eq:kF}
\end{equation}
In all plots shown above, we have taken $k_F a < 0.08$ [using Eq.~\eqref{eq:kF} for $k_F$] which is sufficient for convergence.  For the strong spin-orbit regime, obtaining convergence is more numerically challenging as can be gleaned from the expression in Eq.~\eqref{eq:kF}.  For example, the right panel of Fig.~\ref{fig:StrongSO_FieldScans} ($m\alpha^2/\Delta = 14, L/\ell = 20$) required taking $N_\mathrm{sites}\approx7000$.

\bibliography{Phase_transition}

\end{document}